\documentclass[amsmath,amssymb,prb,floatfix,twocolumn,reprint,longbibliography,superscriptaddress,showpacs,showkeys]{revtex4-1}

\usepackage{graphicx}
\usepackage{dcolumn}
\usepackage{times}
\usepackage{colordvi,epsf}
\usepackage{color}
\usepackage[normalem]{ulem}
\usepackage{amsmath,xfrac,siunitx,bm}

\let\mathitori=\mathrm
\def\mathrm#1{{\mathcode`-=`- \mathitori{#1}}}

\renewcommand{\eqref}[1]{(\ref{#1})}

\newcommand{\ie}{i.e.\@ }

\newcommand{\etal}{\textit{et al.\@ }}

\newcommand{\fleur}{\textsc{Fleur\@}}

\begin{document}

\title{Stability and magnetic properties of Fe double-layers on Ir (111)}
\author{Melanie Dup\'e}
\affiliation{INSPIRE Group, Institute of Physics, Johannes Gutenberg-Universit\"at, 55128 Mainz, Germany}
\author{Stefan Heinze}
\affiliation{Institute of Theoretical Physics and Astrophysics, Christian-Albrechts-Universit\"at zu Kiel,
24098 Kiel, Germany}
\author{Jairo Sinova}
\affiliation{INSPIRE Group, Institute of Physics, Johannes Gutenberg-Universit\"at,
55128 Mainz, Germany}
\affiliation{Institute of Physics ASCR, v.v.i., Cukrovarnicka 10, 162 53 Praha 6 Czech Republic}
\author{Bertrand Dup\'e}
\affiliation{INSPIRE Group, Institute of Physics, Johannes Gutenberg-Universit\"at, 55128 Mainz, Germany}
\affiliation{Institute of Theoretical Physics and Astrophysics, Christian-Albrechts-Universit\"at zu Kiel,
24098 Kiel, Germany}

\date{\today}

\begin{abstract}
We investigate the interplay between the structural reconstruction and the magnetic properties of Fe double-layers on Ir (111)-substrate using first-principles calculations based on density functional theory and mapping of the total energies on an atomistic spin model. We show that, if a second Fe monolayer is deposited on Fe/Ir (111), the stacking may change from hexagonal close-packed to bcc (110)-like accompanied by a reduction of symmetry from trigonal to centered rectangular. Although the bcc-like surface has a lower coordination, we find that this is the structural ground state. This reconstruction has a major impact on the magnetic structure. We investigate in detail the changes in the magnetic exchange interaction, the magnetocrystalline anisotropy, and the Dzyaloshinskii Moriya interaction depending on the stacking sequence of the Fe double-layer. Based on our findings, we suggest a new technique to engineer Dzyaloshinskii Moriya interactions in multilayer systems employing symmetry considerations. The resulting anisotropic Dzyaloshinskii-Moriya interactions may stabilize higher-order skyrmions or antiskyrmions.
\end{abstract}

      
\maketitle

\section{Introduction}

The next generation of high-density and low-energy data storage devices or neuromorphic computing based units will require novel materials and phenomena. 
  Skyrmions in magnetic materials have high potential to meet the demands for these new technologies.\cite{Pinna2017,Prychynenko2017} In condensed matter, magnetic skyrmions were predicted and first studied based on continuous micromagnetic models.\cite{Bogdanov1989,Bogdanov1994} Their existence was confirmed experimentally in bulk and thin-film semiconductors,\cite{Muhlbauer2009a,Yu2010} in metallic multilayers \cite{Moreau-Luchaire2016,Boulle2016} and ultra-thin films.\cite{Heinze2011,Romming2013} The presence and the manipulation of isolated skyrmions in magnetic thin-films and multilayers make them promising for technological applications such as the race-track memory.\cite{Kiselev2011,Fert2013,Sampaio2013,Nagaosa2013}

The presence of isolated skyrmions is attributed to the Dzyaloshinskyi-Moriya interaction (DMI) which occurs where spatial inversion symmetry is broken. \cite{Dzyaloshinskii1957, Moriya1960a} In B20 compounds such as MnSi \cite{Muhlbauer2009a} or Fe$_{0.5}$Co$_{0.5}$Si \cite{Yu2010} this symmetry is broken due to the crystal lattice whereas in the multiferroic Cu$_2$OSeO$_3$\cite{Seki1304201} the polarization breaks inversion symmetry. At surfaces and interfaces, the inversion symmetry is broken due to the interface between different materials.

The DMI originates from spin-orbit coupling (SOC). At metal surfaces and interfaces, the DMI can be understood via the model of Fert and Levy,\cite{Fert1980} which gives a general direction to control the DMI. In ultra-thin films, DMI can be engineered by combining $3d$ transition metals with $4d$ or $5d$ substrates, which provide large SOC.\cite{Belabbes2016a,Herve2017,Dupe2016} In magnetic multilayers, two interfaces can be used to control different magnetic interactions. One interface can be used to tune the magnetic exchange while the other one can generate a large DMI.\cite{Dupe2016} When both interfaces are composed of $5d$ metals, the contribution of each interface can be engineered to obtain a giant DMI.\cite{Moreau-Luchaire2016,Hanke2017}

In ultra-thin films, not only the DMI can be tuned via the interface but also the magnetic exchange interactions, which make them an ideal playground to study magnetism.\cite{Hardrat2009} Among them, the Fe monolayer on Ir (111) has attracted particular attention due to its versatility. If the Fe atoms are adsorbed on the fcc surface sites, the magnetic ground state is a square lattice of skyrmions.\cite{Heinze2011} If Fe is adsorbed in the hcp stacking, the ground state is a hexagonal lattice of skyrmions.\cite{VonBergmann2015} If two monolayers of Fe are deposited on Ir (111), the growth is not epitaxial anymore but results in a complex reconstruction leading to a mixture of fcc, hcp and bcc-stacking of the second Fe-layer characterized by a certain pattern of reconstruction lines.\cite{Hsu2016,Hauptmann2018} These reconstruction lines play a prominent role in the triple layer of Fe on Ir (111) for the writing and deleting of skyrmions by applying an electric field.\cite{Hsu2017} In this system, the surface reconstruction stabilizes skyrmions with an oval shape, which was attributed to an environment anisotropy.\cite{Hagemeister2016a}
Recently, it was found that the symmetry of the interface and thereby the symmetry of the DMI could also determine the type of skyrmions that can be stabilized, i.e. skyrmions or antiskyrmions.\cite{Hoffmann2017} 

In the case of an fcc (100) or an fcc (111) interface, the symmetry of the interface imposes that the DMI has the same sign along each neighboring bond. This configuration favors the presence of skyrmions and explains why Pd/Fe/Ir (111) exhibits isolated skyrmions.\cite{Romming2013,Dupe2014,Simon2014} In the case of a bcc (110) interface, the sign of the DMI may change depending on the nearest neighbor bonds.\cite{Camosi2017,Hoffmann2017} Therefore, antiskyrimons may be more stable, as was illustrated in the case of 2Fe/W (110). \cite{Hoffmann2017} Independently from the symmetry argument, it was also shown that skyrmions and antiskyrmions may coexist in the case of frustrated exchange interaction.\cite{Dupe2014,Leonov2015,Dupe2016,Dupe2016a,Palotas2016,VonMalottki2017} Therefore, an accurate theoretical description of all magnetic interactions is required.

Here, we study the double-layer of Fe on Ir (111) (2Fe/Ir (111)) via density functional theory (DFT) with a particular focus on the different stackings of the two Fe layers of 2Fe/Ir (111). We base our DFT calculations on the experimental observations of certain structural phases.\cite{Hsu2016,Hauptmann2018}
First, a pseudomorphically strained double-layer with fcc-stacking of the surface Fe-layer was identified, which exhibits spin spirals of short periodicity of about 1.2 nm without a preferred propagation direction, indicated by a grainy contrast in the spin-polarized scanning tunneling microscopy (SP-STM) images. 
These areas seem to be prone to defects such as vacancies and substitutional atoms.\cite{Hsu2018}
Second, reconstructed areas were suggested with differently oriented bcc domains separated by reconstruction lines with a characteristic distance of 5.2 nm. 
The reconstruction lines compensate for the lattice mismatch between the Fe layer at the interface, which has a fcc (111)-structure, and the Fe layer at the surface which adopts the bcc (110)-structure.
In the bcc domains spin spirals with periodicities of about 1.9 nm were observed which propagate only along the [100] directions of the bcc-unit cells and thus, in the presence of the reconstruction lines and different  domain orientations, give rise to a characteristic zig-zag shaped herringbone-like magnetic contrast with pearls along the spin spiral propagation directions.

We show that, counterintuitively, the second monolayer does not grow in the fcc or hcp absorption site but in the bcc absorption site. This induces a reduction of the crystal symmetry of 2Fe/Ir (111) which loses the 3-fold rotation axis. We calculate the magnetic exchange interaction for each of the stackings of the Fe second layer and show that the frustration of exchange interaction varies considerably from one stacking to the other. Our DFT-parameterized atomistic spin model includes both intra- and inter-layer exchange interactions. Then, we compute the DMI for each of the stackings and show that the interfacial symmetry does not impose the symmetry of the DMI alone. Finally, we deconvolute the DMI contribution of each of the layers and infer a new method based on symmetry considerations to obtain an anisotropic DMI which may stabilize antiskyrmions or higher order skyrmions.

The paper is organized in two parts. The first part is dedicated to the methodology used to compute the magnetic exchange interaction and the DMI. The second part presents the results regarding the different magnetic ground states of 2Fe/Ir (111) depending on the double-layer stacking.

\section{Model and Computational Methods}

\subsection{Stacking of the Fe double-layers}

We want to study the effect of simple variations of stacking sequences in the Fe double-layer deposited on an Ir (111) surface and how their structural differences influence the stability and magnetic interactions between the Fe-atoms in the double-layer. 
We chose double-layer structures which are derived from the metallic bulk structures fcc, hcp and bcc. 

\begin{figure}[b]
\centering
\includegraphics[width=0.8\columnwidth]{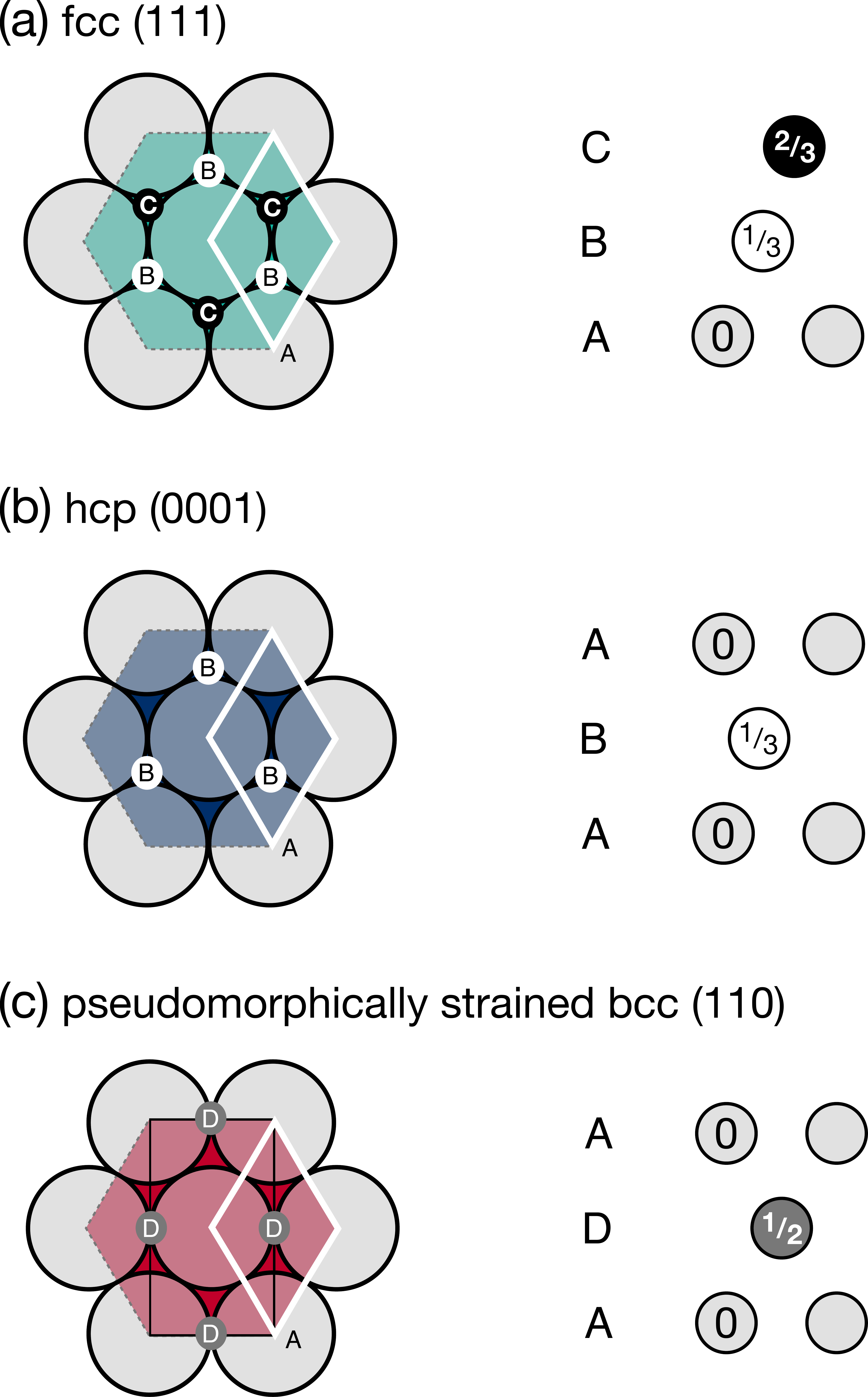}
\caption{(Color online) 
Given are the stacking sequences for the bulk case for structures (a) fcc, (b) hcp and (c) pseudomorphically strained bcc. 
All layers are close-packed in this scenario. }
\label{fig:stackings}
\end{figure}

The fcc structure consists of hexagonal close-packed layers in the (111)-plane, where every atom has six equidistant nearest neighbours within the plane as shown in Fig.~\ref{fig:stackings}(a). These close-packed layers follow an ABC-stacking sequence perpendicular to the plane as shown on the right, where the numbers indicate the x- and y-coordinates of the atoms in the different layers. Each layer occupies a set of hollow sites of the sub-layer.
 
Also the hcp structure in Fig.~\ref{fig:stackings}(b) is formed by hexagonal close-packed layers, which correspond to the (0001)-planes of the hexagonal unit cell. In hcp, the layers follow an ABA-stacking sequence. The hollow sites C remain empty in this case.

The third stacking-type we include in our study, is the pseudomorphically strained bcc (110) structure. In contrast to the fcc and hcp structures, the bcc bulk structure usually does not possess any close-packed crystallographic planes. However, the (110)-plane of the bcc unit cell can be (considerably) strained ($\epsilon_{xx}=-10.7\%$, $\epsilon_{yy}=+9.6\%$) in order to fit the same hexagonal unit cell as the fcc and hcp structures, as indicated in white. The main difference is that the Fe-atoms do not occupy the hollow site positions B or C, but the bridge positions marked D, giving rise to the stacking sequence ADA. Fe atoms in this position have a reduced coordination number as they possess only two nearest neighbours in the plane below instead of three.\\

To characterize the stacking sequences in the Fe double-layer with respect to the fcc-stacked Ir (111) substrate, throughout this paper, we use a modified h-f stacking sequence notation, borrowed from the description of close-packed (bulk) crystal structures. 
In contrast to the original h-f stacking sequence notation, where the symbol always refers to the middle layer of the sequence-triple, i.e. "f" for the B-layer in the ABC-sequence of the fcc structure or "h" for the B-layer in the ABA-sequence of the hcp-structure, our symbols refer to the top layer of the sequence-triple, as we want to characterize surface structures. 
In addition to f and h, we introduce the stacking sequence b$^*$ to indicate a pseudomorpically strained bcc-like top-layer.

\begin{figure}[t]
\centering
\includegraphics[width=1.0\columnwidth]{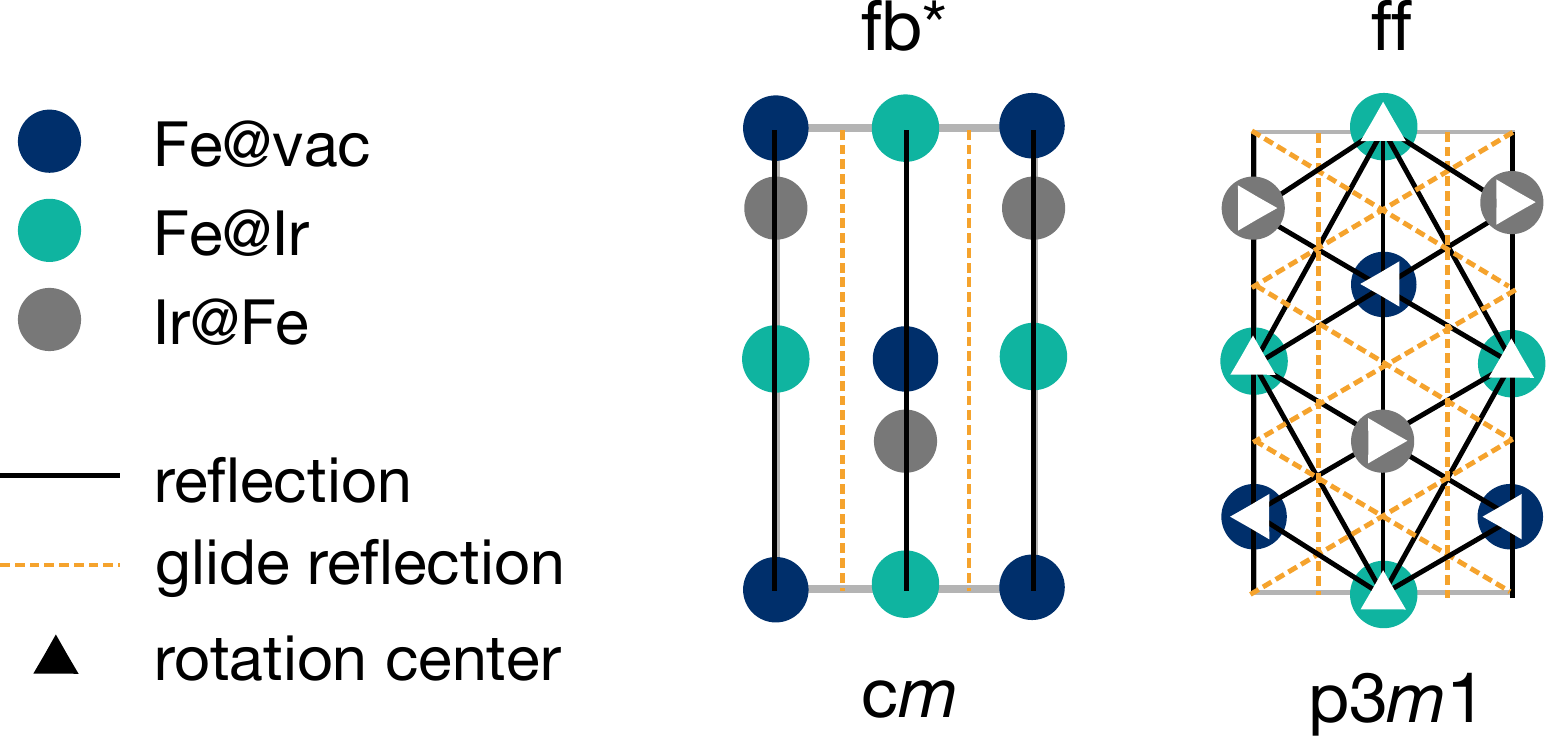}
\caption{(Color online) 
Top views of the atomic configurations of films with stackings fb$^*$ and ff and symmetry elements of their respective plane groups c$m$ and p$3m1$. Shown are the atoms of the three outmost layers, only.}
\label{fig_symmetries2}
\end{figure}  

In Tab.~\ref{tab:stacking-nomenclature} we give an overview of the stacking sequences that we have studied. 
The first symbol indicates the stacking sequence of the Fe atom at the interface Fe@Ir and the second symbol the stacking sequence of the Fe atom at the surface Fe@vac. 
For Fe@Ir only stackings f and h were considered, while for Fe@vac also b$^*$ was taken into account.
Besides the stacking sequence in ABC-notation also the coordination numbers (CN) of the two Fe-atoms are given, demonstrating that coordination numbers are reduced by one, if the top layer adopts the bcc-like structure.
Two different symmetries result in the close-packed structures ff, fh, hf and hh, we find the trigonal plane group (PG) p$3m1$ and in the bcc-like structures fb$^*$ and hb$^*$ the centered rectangular/rhombic plane group c$m$. Figure~\ref{fig_symmetries2} illustrates these symmetries. In an isolated double-layer, the symmetry is higher in the bcc-like stackings (c$2mm$ instead of c$m$), but unchanged in the others.

\begin{table}[b]
\centering
\begin{ruledtabular}
\caption{\label{tab:stacking-nomenclature}Overview of the double-layer structures: 
Given are the names and stacking sequences in ABC-notation, where the stacking of the substrate is given in parentheses. 
Also indicated are the coordination numbers (CN) of the Fe atoms in the two layers of the double-layer, as well as the plane groups (PG) of the isolated double layer and the full film.}
\begin{tabular*}{\hsize}{cccccc}
          & Stacking  & CN        & CN          & PG      & PG \\
Name & sequence & Fe@Ir & Fe@vac &  double-layer  & film\\
 \hline
ff & (ABC) AB & 12 & 9  & p$3m1$ & p$3m1$\\
fh & (ABC) AC & 12 & 9  & p$3m1$ & p$3m1$\\
fb$^*$ & (ABC) AD & 11 & 8 & c$2mm$ & c$m$\\
\hline
hf & (ABC) BA & 12 & 9 & p$3m1$ &  p$3m1$\\
hh & (ABC) BC & 12 & 9 & p$3m1$ & p$3m1$\\
hb$^*$ & (ABC) BD & 11 & 8 & c$2mm$ & c$m$\\    
\end{tabular*}
\end{ruledtabular}
\end{table}

\subsection{Stability of the stackings}

We study via density functional theory (DFT) calculations the energies and magnetic interactions of the six different structural stackings presented in Table~\ref{tab:stacking-nomenclature}. We have used the \fleur\ ab initio package.\cite{FLEUR} The \fleur\ code utilizes the full potential linearized augmented plane wave approach (FLAPW),\cite{Krak1979,Wimmer1981,Weinert1982} which ranks among the most accurate electronic structure techniques. Especially, \fleur\ allows us to study complex magnetic states at interfaces such as non-collinear magnetic states,\cite{Bode2007} skyrmion lattice ground states \cite{Heinze2011} and the presence of isolated topologically protected states in bilayers such as skyrmions or antiskyrmions.\cite{Dupe2014,Dupe2016a,Dupe2016,Hoffmann2017}

To optimally describe the geometry of the Fe/Ir interface, a mixed exchange correlation functional was employed.\cite{DeSantis2007} 
This mixed functional applies the generalized gradient approximation in the parameterization of Perdew \etal (GGA)\cite{Perdew1992} to the interstitial region and to the muffin-tin spheres of the Fe atoms, whereas in the Ir atoms' muffin-tin (MT) spheres the local density approximation (LDA)\cite{Vosko1980} is applied. 
This method has been shown to capture the magnetic and structural properties of the $3d$ elements on $5d$ substrates.\cite{Baud2006,Mokrousov2009,Menzel2012}

The muffin-tin radii were set to 2.23 bohr (1.18 \AA) for Fe and 2.31 bohr (1.22 \AA) for Ir. 
We chose a plane-wave cut-off $k_\text{max}$ of 4.0 bohr$^{-1}$ and a mesh of 256 $k$-points within the irreducible part of the first Brillouin zone.

Structural relaxations were performed using a symmetric film consisting of 11 layers of Ir and two layers of ferromagnetic (FM) Fe on the top and bottom of the Ir slab, slightly different from the slab shown in Fig.~\ref{fig:slab}. The equilibrium hexagonal lattice parameter of the Ir substrate of $a_\text{Ir}=5.10$~bohr ($2.70$~\AA) was used.
The atoms of the outmost three layers, i.e. Fe@vac, Fe@Ir and Ir@Fe, were allowed to relax along the z-direction until forces were smaller than 0.001 Hartree per bohr (0.04 eV/\AA). 
Table~\ref{tab:distances} provides the resulting inter-layer distances. 
$d_\text{Ir-Ir}$ is with 2.23--2.25~\AA\ larger than the distance in the bulk material of 2.20~\AA, it shows little variation for the different stackings but tends to be smaller for the h$x$-stackings. 
$d_\text{Fe-Ir}$ depends weakly and $d_\text{Fe-Fe}$ depends strongly on the stacking of the Fe@vac atoms. 
The values for $d_\text{Fe-Ir}$ are in the range between 2.09 and 2.12~\AA.
While $d_\text{Fe-Fe}$ is with 2.08~\AA\ largest for fh- and hh-stacking, it is considerably smaller in ff, hf, fb$^*$ and hb$^*$ with 2.01--2.02~\AA.

\begin{table}[tb]
\centering
\begin{ruledtabular}
\caption{\label{tab:distances}Inter-layer distances for the outmost three layers in the relaxed structures. Values are given in \AA.}
\begin{tabular*}{\hsize}{ccccccc}
 & ff & fh & fb$^*$ & hf & hh & hb$^*$ \\
 \hline
$d_\text{Fe-Fe}$ & 2.02 &  2.08 & 2.01 & 2.01 & 2.08 & 2.02\\
$d_\text{Fe-Ir}$  & 2.09 &  2.11 & 2.12 & 2.09 & 2.11 & 2.11\\
$d_\text{Ir-Ir}$    & 2.25 &  2.25 & 2.24 & 2.23 & 2.23 & 2.23\\
\end{tabular*}
\end{ruledtabular}
\end{table}

\subsection{Symmetry aspects of the spin spirals}

A powerful way of describing and understanding the underlying mechanisms and magnetic interactions leading to non-collinear magnetic structures in ultrathin films is to study the energy dispersion of spin spirals.\cite{Sandratskii1991,Kurz2004}

Homogeneous spin spirals possess static periodic structures, which can be incommensurate with the chemical unit cell of the crystal lattice. 
Therefore, we describe them by wave vectors $\mathbf{q}$ in reciprocal space so that the magnetic moment $\bold{M}$ at site $\bold{r}$ is given by:

\begin{eqnarray}
\bold{M} \left( \bold{r} \right)=M \left( \begin{array}{c}
 \mathrm{sin} \left( \bold{qr}\right) \\
 \mathrm{cos} \left( \bold{qr}\right) \\
 0
 \end{array}
 \right),
\end{eqnarray}

\begin{figure}[tb]
\centering
\includegraphics[width=1.0\columnwidth]{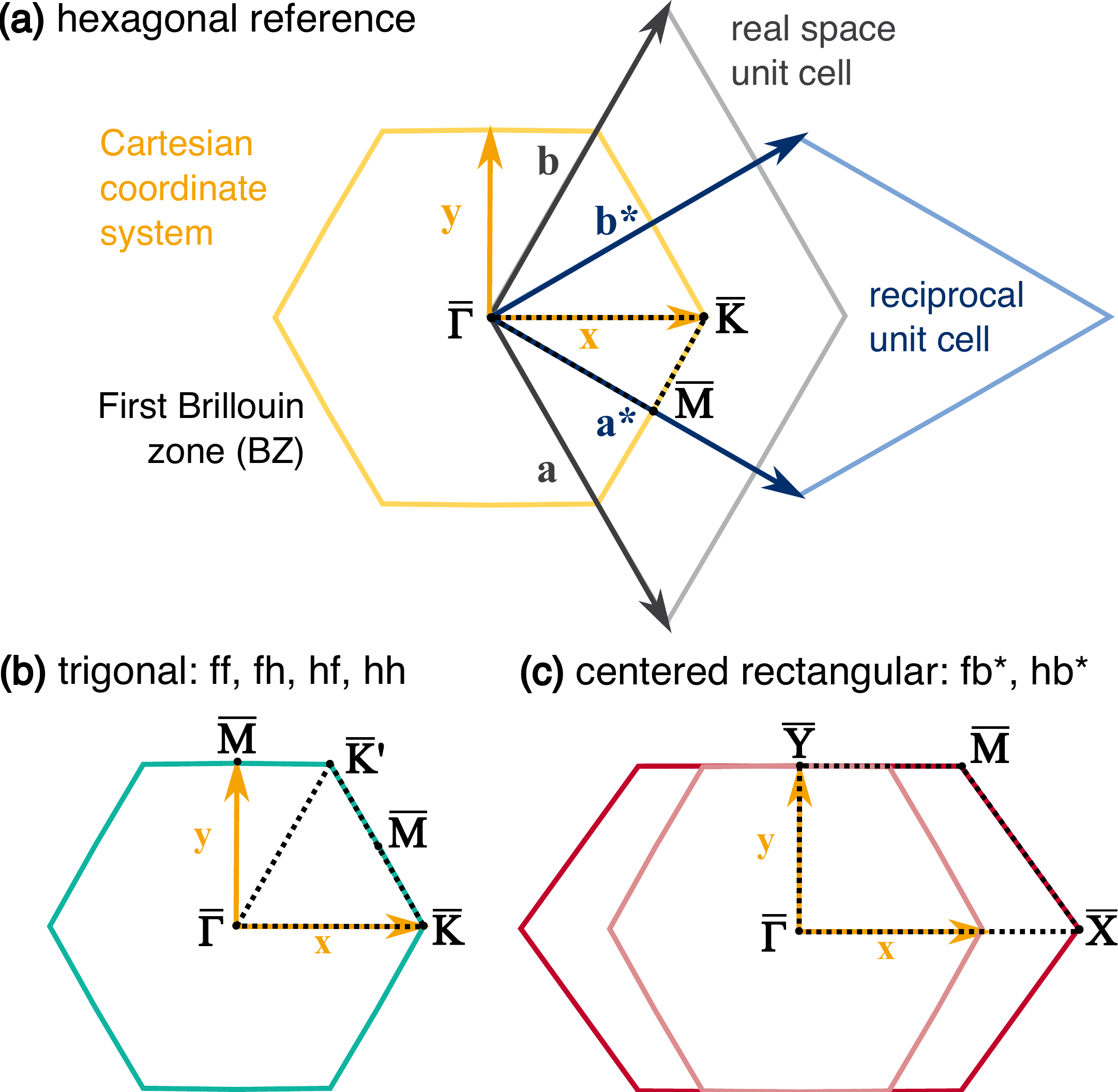}
\caption{(Color online) 
Brillouin zones of the double-layer stackings. 
In (a) the relationships between real space (grey) and reciprocal (blue) unit cells including the first Brillouion zone (yellow) for the hexagonal reference structure is given. 
For simplicity, the vectors of the real space and reciprocal space were chosen to have the same length. The dashed black triangle indicates the irreducible part of the first Brillouin zone.
Also shown is the cartesian coordinate system within the first Brillouin zone (yellow), which is used for the calculation of magnetic interaction energies. 
The high symmetry $q$-points $\bar{\Gamma}$,  $\bar{\mathrm{K}}$ and  $\bar{\mathrm{M}}$ are indicated as well as the irreducible part of the Brillouin zone (dotted line). 
In the double-layers, the hexagonal symmetry is reduced to trigonal and centered rectangular. 
In (b) the trigonal setting is shown where half of the points $\bar{\mathrm{K}}$ are lost, indicated by the additional $\bar{\mathrm{K}}'$ and a doubling of the size of the irreducible Brillouin zone. 
In the centered rectangular systems, given in (c), the Brillouin zone changes its shape and size (the hexagonal one is shown for comparison) and new special $k$-points $\bar{\mathrm{X}}$, $\bar{\mathrm{Y}}$ and $\bar{\mathrm{M}}$ result.}
\label{fig:reciprocalspace}
\end{figure}

This spin spiral propagation vector $\mathbf{q}$ can be chosen arbitrarily in the irreducible Brillouin zone. In Fig.~\ref{fig:reciprocalspace}, we present the irreducible Brillouin zones of the different double-layer stackings. In Fig.~\ref{fig:reciprocalspace}(a), we show how the unit cells of the real space (spanned by $\mathbf{a}$ and $\mathbf{b}$, in grey) and the reciprocal space (spanned by $\mathbf{a}^*$ and $\mathbf{b}^*$, in blue) are related to each other for a hexagonal lattice (the length of the vectors is arbitrary here). In internal coordinates, the high-symmetry points $\bar\Gamma=(0,0)$, $\bar{\mathrm{M}}=(\tfrac{1}{2},0)$ and $\bar{\mathrm{K}}=(\tfrac{1}{3},\tfrac{1}{3})$ within the first Brillouin zone (BZ, shown in yellow) are indicated as well, along with the irreducible part of the BZ (dotted line).
To extract the exchange and the DM interactions, we employ the cartesian coordinate system within the BZ, which is spanned by $\mathbf{x}$ and $\mathbf{y}$ (in yellow).

As we learned before (see Tab.~\ref{tab:stacking-nomenclature}), the Fe double-layer does not possess the six-fold rotation axis of the hexagonal symmetry. Instead, depending on the stacking sequence, the symmetry is reduced to trigonal (plane group p$3m1$) in the hexagonal close-packed stackings ff, fh, hf, and hh or to centered rectangular (plane group c$2mm$) in the bcc-like stackings fb$^*$ and hb$^*$.  Therefore, some of the high-symmetry points are lost in the close-packed stackings or become fully obsolete in the bcc-like stackings, as indicated in Fig.~\ref{fig:reciprocalspace}(b) and (c).

In the case of trigonal symmetry, the $\bar{\mathrm{K}}$-points are not equivalent anymore as shown by the additional $\bar{\mathrm{K}}'$ along $\vec{b}$. The $\bar{\mathrm{M}}$-points remain unchanged. Therefore, the irreducible BZ has doubled in size as compared to the case shown in Fig.~\ref{fig:reciprocalspace}(a). In the centered rectangular symmetry, the BZ changes its shape and size. The former $\bar{\mathrm{M}}$-point along $\mathbf{y}$ becomes the $\bar{\mathrm{Y}}$-point at $(0,\tfrac{1}{\sqrt{3}})$ (in cartesian coordinates) and the $\bar{\mathrm{K}}$-point becomes obsolete. 
A new high-symmetry $\bar{\mathrm{X}}$-point results at $(1,0)$ and a new $\bar{\mathrm{M}}$-point at $(\tfrac{1}{\sqrt{3}},\tfrac{1}{\sqrt{3}})$.\\

As $\bar\Gamma\text{-}\bar{\mathrm{M}}$ and $\bar\Gamma\text{-}\bar{\mathrm{X}}$ along $\mathbf{x}$ as well as $\bar\Gamma\text{-}\bar{\mathrm{K}}$ and $\bar\Gamma\text{-}\bar{\mathrm{Y}}$ along $\mathbf{y}$ are high-symmetry lines in both the trigonal and centered rectangular symmetries, we utilize these throughout this study to compare the magnetic interactions in the different stackings.

The spin spiral propagation vector $\mathbf{q}$ determines the propagation direction of the spin spiral and the periodicity length or, in other words, the angle between the neighbouring magnetic moments. A spin spiral with $\mathbf{q}=\Gamma$ describes the ferromagnetic state. At  $\mathbf{q}=\bar{\mathrm{M}}$ the row-wise antiferromagnetic state is characterized by an angle of $180^{\circ}$ between neighbouring magnetic moments and a periodicity length of $2a$. $\mathbf{q}=\bar{\mathrm{K}}$ on the other hand characterizes the N\'eel state with an angle of $120^{\circ}$ and a periodicity length of $3a$.

\begin{figure}[tb]
\centering
\includegraphics[width=1.0\columnwidth]{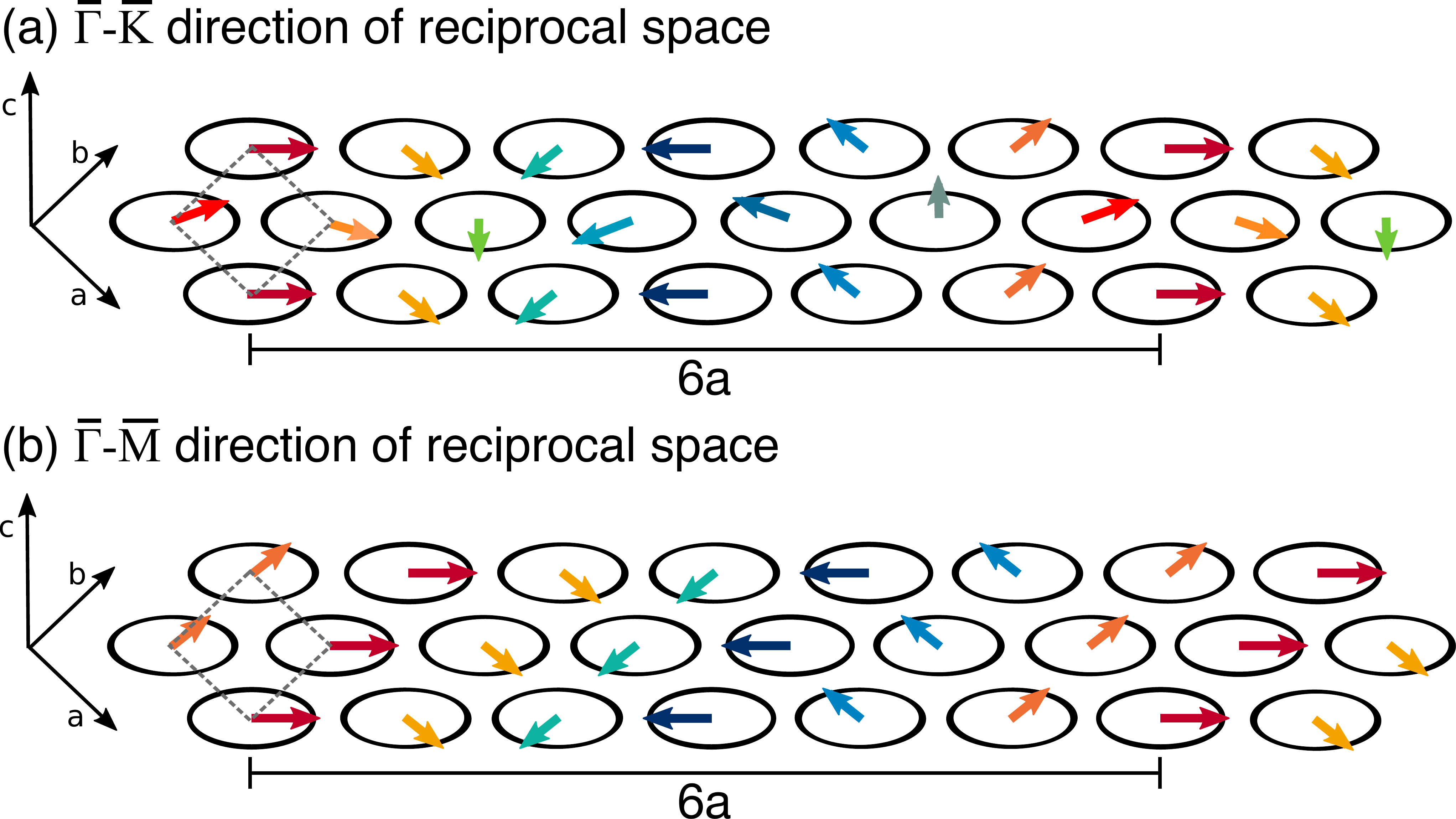}
\caption{(Color online) 
Schematic pictures of flat spin spirals on the hexagonal lattice in real space corresponding to $q=\frac{1}{6}\frac{2\pi}{a}$ propagating along (a) $\bar{\Gamma}\text{-}\bar{\mathrm{K}}$ direction with the wave vector  $(q,0,0)$ and (b) $\bar{\Gamma}\text{-}\bar{\textrm{M}}$ direction of reciprocal space with the wave vector $(0,q,0)$.}
\label{fig:flatspinspiral}
\end{figure}

In Fig.~\ref{fig:flatspinspiral} two examples are shown corresponding to $q=\frac{1}{6}\frac{2\pi}{a}$ propagating along the two directions $\bar{\Gamma}\text{-}\bar{\mathrm{K}}$ with the wave vector $(q,0,0)$ and $\bar{\Gamma}\text{-}\bar{\textrm{M}}$ with the wave vector $(0,q,0)$.

\subsection{Magnetic interactions from DFT calculations}

\subsubsection{Magnetic exchange interaction}

\begin{figure}[tb]
\centering
\includegraphics[width=0.45\columnwidth]{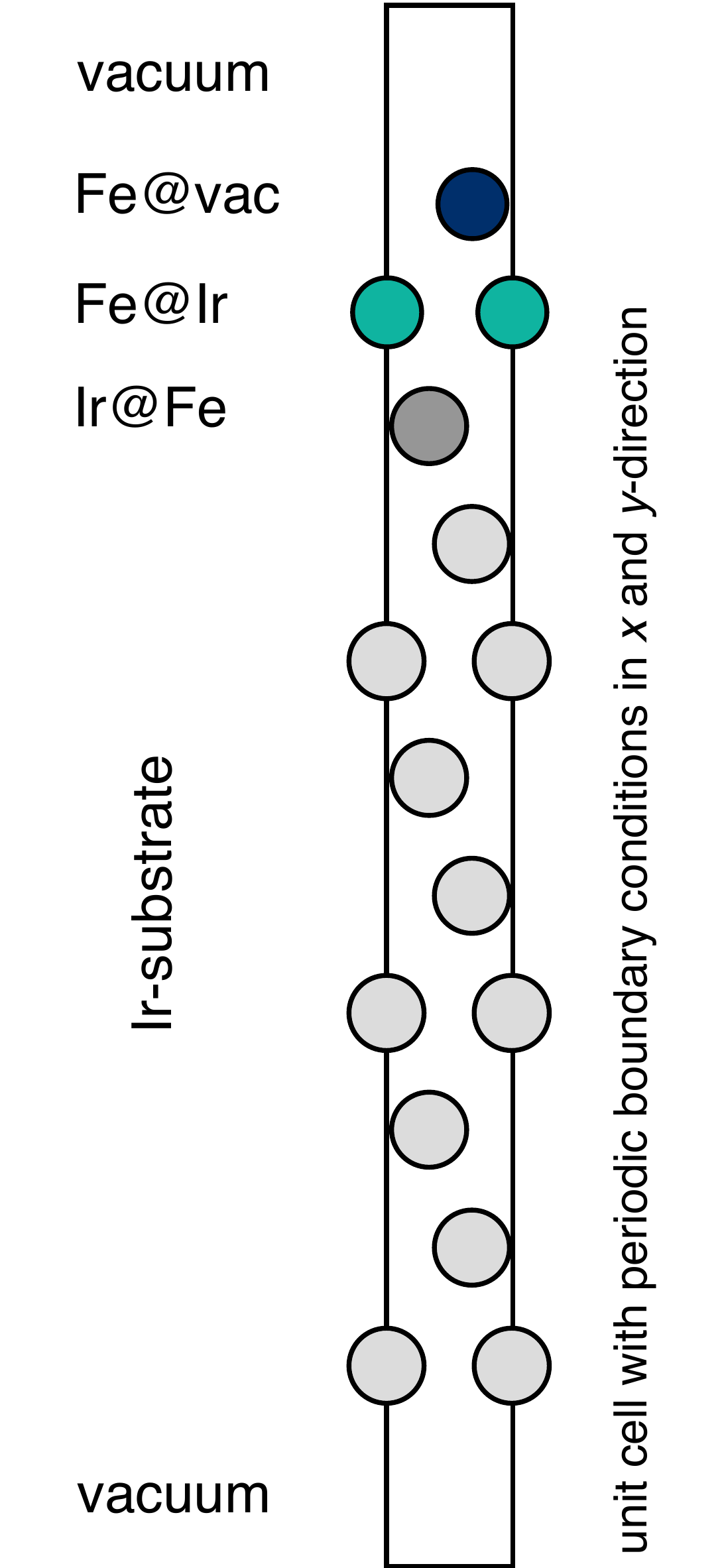}
\caption{(Color online) 
Asymmetric film geometry as it was used in the calculations of the magnetic properties. 
Shown is the ff-stacking of the Fe double-layer.}
\label{fig:slab}
\end{figure}

To study the magnetic properties of these films, we calculate the total energy of flat spin spirals as a function of the angle between neighboring magnetic moments via the generalized Bloch theorem.\cite{Kurz2004} It allows the calculation of homogeneous spin spirals which are incommensurate with the chemical unit cell of the crystal lattice. Here, we only consider flat spin spirals which are propagating in the xy-plane and are described by the propagation vector $\mathbf{q}$. To minimize the computational cost, we have used an asymmetric slab consisting of two iron atoms on nine layers of iridium substrate as shown in Fig.~\ref{fig:slab}. In these calculations, the spin spiral is propagating in both the iron and the iridium layers unless stated otherwise. We use a $k_\text{max}=4.0$~bohr$^{-1}$ and 1936~$k$-points in the full BZ.

\subsubsection{Spin-orbit coupling contributions}

SOC contributes to the  magnetocrystalline anisotropy energy (MAE) and to the DMI. The MAE contribution is obtained from calculations of collinear spin structures. Whereas the DMI energy contribution is accessed via calculations of spin spirals.

The MAE can be obtained by evaluating the energy contribution of the SOC in the collinear Fe double-layers with all magnetic moments oriented parallel along the three cartesian axes $x$, $y$ and $z$. 
This SOC contribution is calculated by performing self-consistent scalar-relativistic calculations, which require an increased accuracy, therefore we use a $k_\mathrm{max}=4.3$~bohr$^{-1}$ and 1936 $k$-points. To study the stability of the MAE with respect to the number of Ir layer, we have successively turned off the SOC contribution in the muffin tin of the Ir atoms. In that case, we have converged self-consistently the charge density when the quantization axis was applied in the $z$-direction and applied the magnetic force theorem to evaluate the energy when quantization axis was applied in the $x$- and $y$-direction.

For the non-collinear case of a flat spin spiral, the SOC can be included via first-order perturbation theory.\cite{Heide2009} In that case, the band energies are corrected via the SOC Hamiltonian
\begin{equation}
H_{\mathrm{SOC}}=\sum_i \xi_i \bm{\mathrm{\sigma}} \cdot \bm{\mathrm{L}}_i,
\end{equation}
where $\xi_i$ is the SOC strength at site $i$, $\bm{\mathrm{\sigma}}$ is the Pauli matrix and $\bm{\mathrm{L}}_i$ is the orbital momentum operator at site~$i$. $H_{\mathrm{SOC}}$ describes the odd part of the magnetic exchange tensor which can be interpreted as the DMI, which stabilizes a left or a right rotating spin-spiral.
Within \fleur, the FLAPW basis set provides a natural framework for the atomic decomposition of the SOC. 
The atomically resolved SOC contribution allows the determination of the layer-dependent contributions to the DMI. If the magnetic moments in the MT of the layer $i$ are kept in a FM state perpendicular to the rotation plane of the spin spiral in the other layers then the SOC contribution of this layer will be reduced to zero. The DMI can be attributed to the SOC contributions of the remaining layers. For these calculations, we have used $k_\mathrm{max}=4.3$~bohr$^{-1}$ and 1936 $k$-points. \\

\subsection{Extended Heisenberg model}

We use an extended Heisenberg model to analyze the energy dispersion curves $E(\bm{q})$ of the different stacking models. This model involves the magnetic exchange interactions up to the fifth nearest neighbors and the Dzyaloshinksii-Moriya interactions.

\subsubsection{Heisenberg exchange interaction}
\label{Exch_mag}

The exchange interactions in the magnetic Fe double-layer can be separated into two contributions:\cite{Dupe2016} the intra-layer interactions within the layers parallel to the film (labeled Fe@vac and Fe@Ir) and the inter-layer interactions between these two layers perpendicular to the film. This decomposition results in the spin Hamiltonian
\begin{equation}
H=H^{\parallel,\mathrm{Fe@vac}}+H^{\parallel,\mathrm{Fe@Ir}}+H^\perp.
\end{equation}

Both the intra- and the inter-layer interaction Hamiltonian are expressed as
\begin{equation}
H^{\parallel,\perp}=-\sum_{ij}J_{ij}^{\parallel,\perp}(\mathbf{m}_i \cdot \mathbf{m}_j),
\label{eq:Ham}
\end{equation}
where the sum runs over sites within both Fe-layers.
This Hamiltonian may be expressed as a series of Cosines by inserting the magnetization of a homogeneous spin spiral as

\begin{equation}
H=-\sum_\delta J_\delta \sum_i \cos(\mathbf{q} \cdot \mathbf{R}_{\delta i}),
\end{equation}
where $\mathbf{R}_{\delta i}$ is the position of the atom $i$ in the shell $\delta$ and $\mathbf{q}$ is the propagation vector of the spin spiral in units of $\tfrac{2\pi}{a}$.\\
 
In all structures presented in this study, both Fe layers adopt a hexagonal close-packed structure. Therefore $H^{\parallel}$ remains unchanged as compared to previous works.\cite{Dupe2016} In contrast, $H^{\perp}$ depends on the double layer stacking. In the case of the close-packed stackings ff, fh, hf and hh, there are three next nearest neighbors in the adjacent plane at positions $(\tfrac{a}{2},\tfrac{a}{2\sqrt{3}})$, $(-\tfrac{a}{2},\tfrac{a}{2\sqrt{3}})$ and $(0,-\tfrac{a}{\sqrt{3}})$. The resulting equations for the inter-layer interactions can be found in the Supplemental Material of Ref[\onlinecite{Dupe2016}].\\

For the bcc-like stackings fb$^*$ and hb$^*$, there are two next nearest neighbor atoms at positions $(\tfrac{a}{2},\tfrac{a}{2})$ and $(-\tfrac{a}{2},\tfrac{a}{2})$. This gives for the first four neighbor shells in the adjacent plane the following expressions in cartesian coordinates $H_{nn}^\perp(q_x,q_y)$:

\begin{align}
H_1^\perp&=2 J_1^\perp \Big[\cos\Big(\frac{aq_x}{2}\Big) - 1\Big] \label{H_1} \\
H_2^\perp&=2 J_2^\perp \Big[\cos\Big(\frac{\sqrt{3}aq_y}{2}\Big) - 1\Big] \label{H_2} \\
H_3^\perp&=4 J_3^\perp \Big[\cos(aq_x)  \Big(\frac{\sqrt{3}aq_y}{2}\Big) - 1\Big] \\
H_4^\perp&=2 J_4^\perp \Big[\cos\Big(\frac{3aq_x}{2}\Big)-1\Big]
\end{align}

Thus, the energy dispersions obtained for three different spin spirals in the Fe-layers can be fitted to the spin Hamiltonian \eqref{eq:Ham} in order to obtain the exchange interaction coefficients $J_{ij}^{\parallel,\mathrm{Fe@vac}}$, $J_{ij}^{\parallel,\mathrm{Fe@Ir}}$ and $J_{ij}^\perp$ up to the fifth and fourth neighbor shell, respectively. $J_{ij}^{\parallel,\mathrm{Fe@vac}}$ and $J_{ij}^{\parallel,\mathrm{Fe@Ir}}$ are obtained when the spin spiral only propagates in the $\mathrm{Fe@vac}$ or $\mathrm{Fe@Ir}$ layer, respectively. $J_{ij}^\perp$ is obtained when the spin spiral propagates in both Fe layers simultaneously. \\
In addition, we provide effective exchange coefficients $J_\mathrm{eff}$ resulting from the first nearest neighbor exchange interaction obtained in the range $|\bm{q}| < 0.1\times(\tfrac{2\pi}{a})$, i.e. in the vicinity of the $\bar{\Gamma}$-point.\cite{Dupe2016,VonMalottki2017}

\subsubsection{Dzyaloshinskii-Moriya interaction}

The DMI arises when SOC occurs in a system with broken inversion symmetry such as a surface or an interface. 
This interaction favours a perpendicular orientation between neighboring spins instead of the parallel or anti-parallel orientation favoured by Heisenberg exchange.
Its effect is to favour cycloidal spin spirals and a certain rotational sense, thus, it determines whether a spin spiral rotates clockwise or counter-clockwise.
The Hamiltonian can be written as
\begin{equation}
H_\mathrm{DM}=-\sum_{ij} \mathbf{D}_{ij}\cdot\big(\mathbf{m}_i\times\mathbf{m}_j\big),\label{eq:DMI}
\end{equation}
where the sum runs over sites within both Fe-layers.

We consider contributions to the DMI up to the third neighbor shell and also provide effective DMI coefficients $D_\text{eff}$ resulting from the next nearest neighbor interaction as obtained close to the $\bar{\Gamma}$-point ($q \in [-0.1,0.1]$). \\

\section{Results and Discussion}

This section is organized as follows.
We first investigate and discuss the stability of the different stackings of the Fe double-layer. We consider the low temperature case (A) as well as a high temperature case (B), where the effect of thermal expansion of the substrate is taken into account. For completeness, we also study the kinetic stability of the stackings with respect to the ground state structure (C).
Afterwards we provide an in-depth investigation of the magnetic interactions playing a decisive role in this material system (D), i.e. Heisenberg exchange (D 1), magnetocrystalline anisotropy (D 2) and the Dzyaloshinskii Moriya interaction (D 3).
In order to understand the obtained relative stabilities and magnitudes of magnetic interactions we present densities of states (E).
Finally, after having determined all energy contributions to the magnetic texture resulting from the different stackings in the Fe double-layer, we can discuss the occurrence of spin spirals in the studied system and compare our findings with experiments.

\subsection{Thermodynamic stability of stackings}

\begin{figure}[tb]
\centering
\includegraphics[width=0.9\columnwidth]{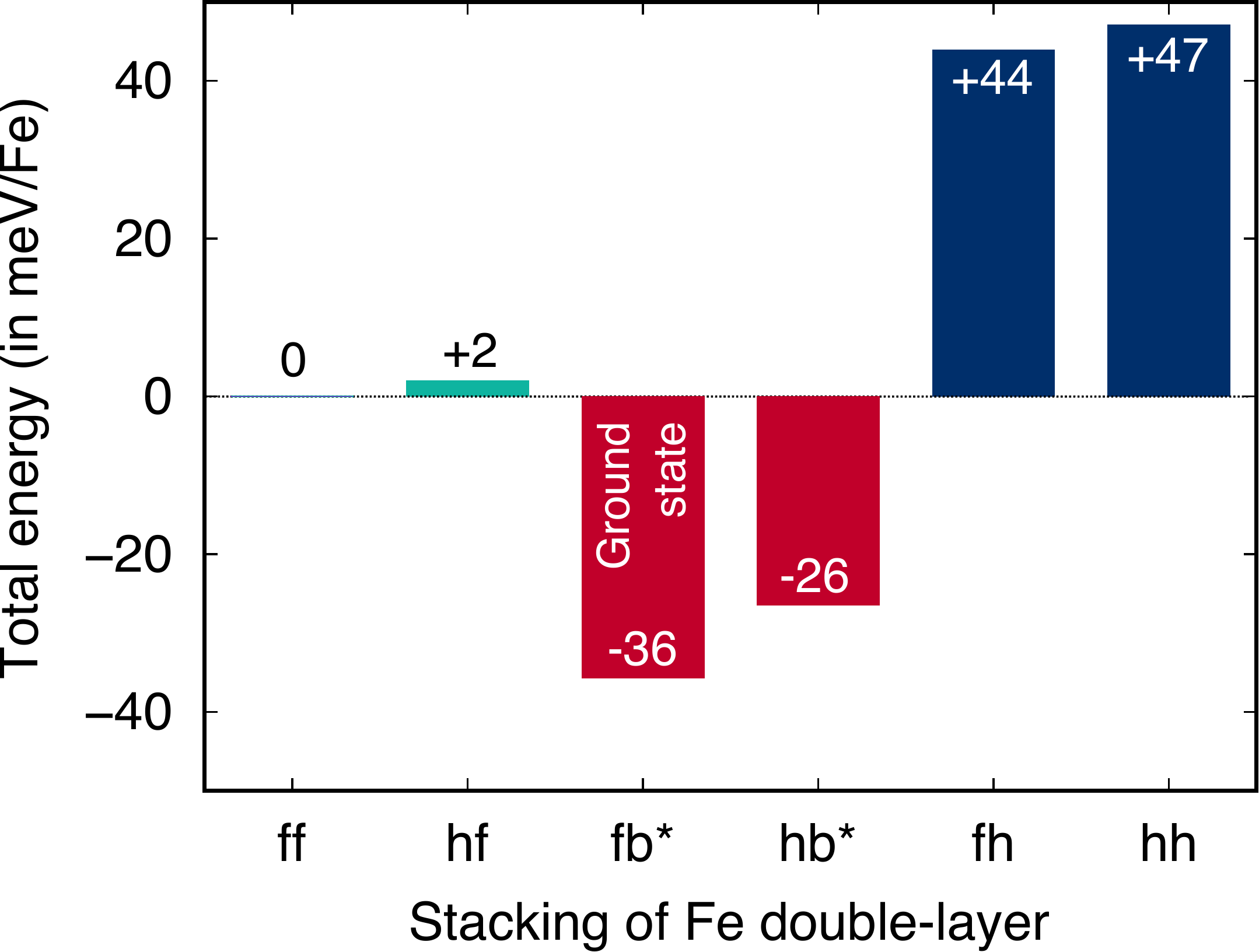}
\caption{(Color online) 
Total energies of the different stackings relative to the energy of the ff-stacking. 
Decisive for the stability of the double layer is the top-layer configuration.}
\label{fig:stability}
\end{figure}

We start by presenting the stability of the different double-layer stackings given in Tab.~\ref{tab:stacking-nomenclature}. 
In Fig.~\ref{fig:stability} the total energies of these different structures are given relative to the energy of the ff-stacking. 
We find three groups of stackings in terms of stability, the $x$f-stackings with energies around zero meV per Fe atom (where $x$ is f, h), the $x$h-stackings with higher energies at about 45 meV per Fe atom and the more stable $x$b*-stackings at about $-30$ meV per Fe-atom, of which fb$^*$ gives the ground state of the Fe double-layer on Ir (111).

This result can be considered surprising in two ways. 
Firstly, the Fe@vac-layer uniquely determines the stability of the double layer. 
Secondly, the low-symmetry structures $x$b*  are more stable than the close-packed stackings $x$h or $x$f. 
It seems that two monolayers of Fe are sufficient to give rise to bulk-like properties (i.e. the FM bcc-ground state structure) in the ultra thin-film, although it is extremely strained: $\epsilon_\mathrm{xx}=-4.6\%$ and $\epsilon_\mathrm{yy}=+17.0\%$ with respect to the calculated lattice constant of bcc iron (compare $a=2.83$~\AA\ of Fe in cubic unit cell vs. $a=2.70$~\AA\ of Ir in hexagonal unit cell). There is no explanation so far for the experimental finding that both fcc and bcc-top layers can be found to coexist,\cite{Hsu2016}  according to our calculations only fb$^*$ and hb$^*$ stackings should be observed in the Fe double-layer.

\subsection{Effect of epitaxial strain}

\begin{figure}[tb]
\centering
\includegraphics[width=0.9\columnwidth]{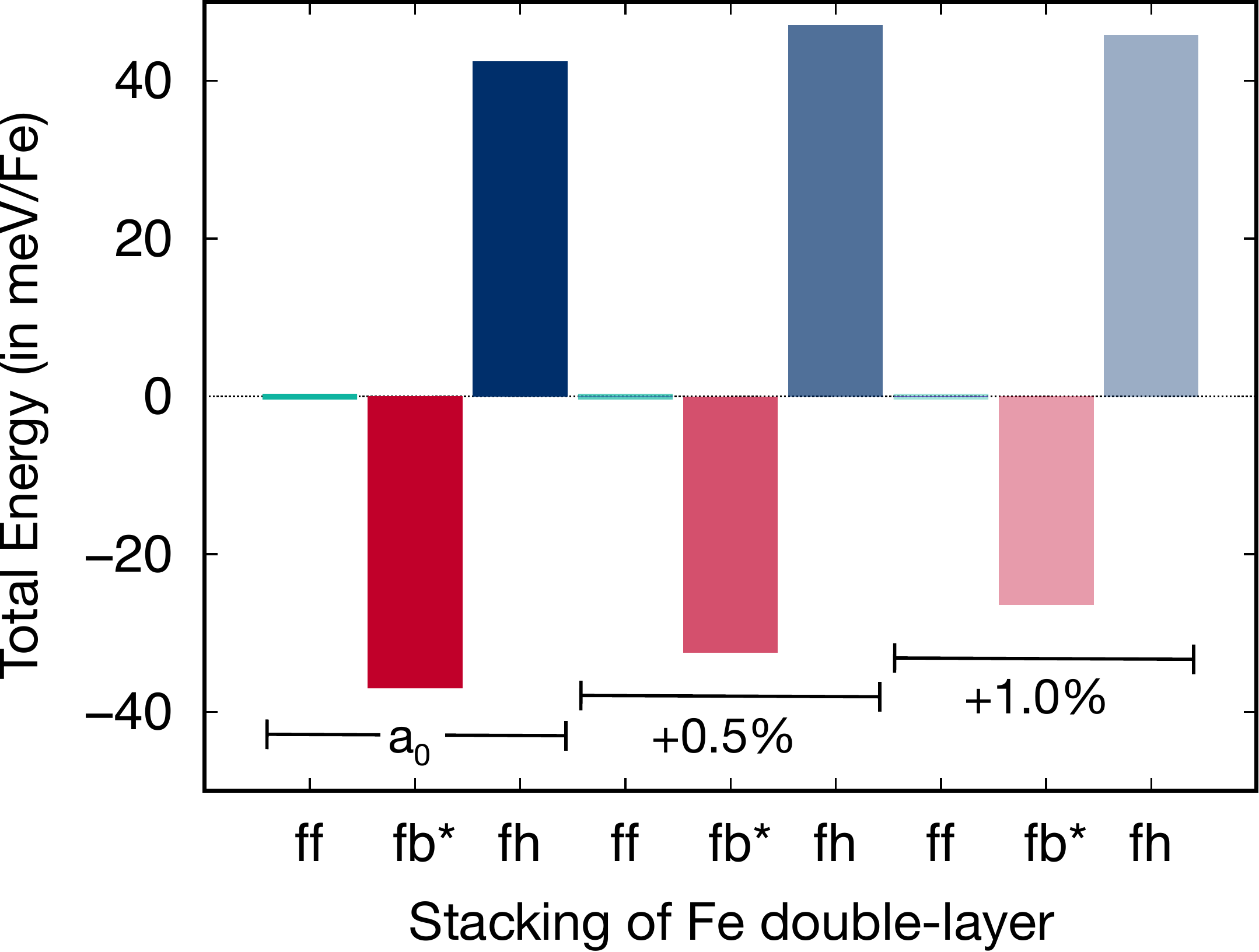}
\caption{(Color online) 
Total energies of the different stackings relative to the energy of the ff-stacking with different epitaxial strains. 
Epitaxial strain does not alter the relative stabilities of the different double-layer stackings.}
\label{fig:strain}
\end{figure}

As the Fe double-layers are grown at elevated temperatures of about 700 Kelvin, thermal expansion of the Ir substrate might affect the relative stabilities of the stackings. 
In order to investigate the effect of an increased epitaxial strain due to thermal expansion of the substrate we studied the f$x$-stackings at different in-plane lattice-constants expanded by 0.5 and 1.0\% as compared to the DFT equilibrium value of $a_0=2.70$~\AA. 
Iridium has an expanded lattice constant by about 0.34\% at 600 K as compared to the value at temperatures close to 0 K,\cite{Arblaster2010} while the DFT value is 0.45\% smaller than the experimental 0 K-value. 
Thus, our range is sufficiently large to cover for both the DFT-underestimation and the thermal expansion effect. The total energies are shown in Fig.~\ref{fig:strain}. 
All values are given relative to the energy of the ff-stacking at the same lattice constant.
 
The energy difference between ff- and fb$^*$-stackings slightly decreases by 10 meV from the equilibrium lattice constant $a_0$ to +1.0\%  epitaxial strain, while the energy of the fh-stacking stays almost constant at a value of about 45 meV per Fe-atom higher than the ff-stacking.
Therefore, epitaxial strain alone cannot change the stability hierarchy of the stackings we studied.

\subsection{Kinetic stability of the bcc-like stacking}

Since epitaxial strain cannot explain the presence of ff-stacked areas in experimentally investigated Fe double-layers, we turn now to the kinetic stabilization of the hexagonal-close packed stackings, \ie if there are any energy barriers between those and the bcc-like stackings. 

We envision that the growth of the second layer could start at the hollow sites, i.e. positions B or C of the close-packed structures with three next neighbors in the Fe@Ir-layer (see Fig.~\ref{fig:stackings}), in contrast to the energetically unfavored D-positions with only two next neighbors characteristic of the bcc-like surface structure.  An energy barrier between the close-packed structures and the bcc-like structure would then explain why larger islands cannot transform into the ground state fb$^*$ but stay in the metastable states ff or fh. Thus, the structure would be determined by growth and kinetic considerations although thermodynamics favor another stacking. 

\begin{figure}[tb]
\centering
\includegraphics[width=0.95\columnwidth]{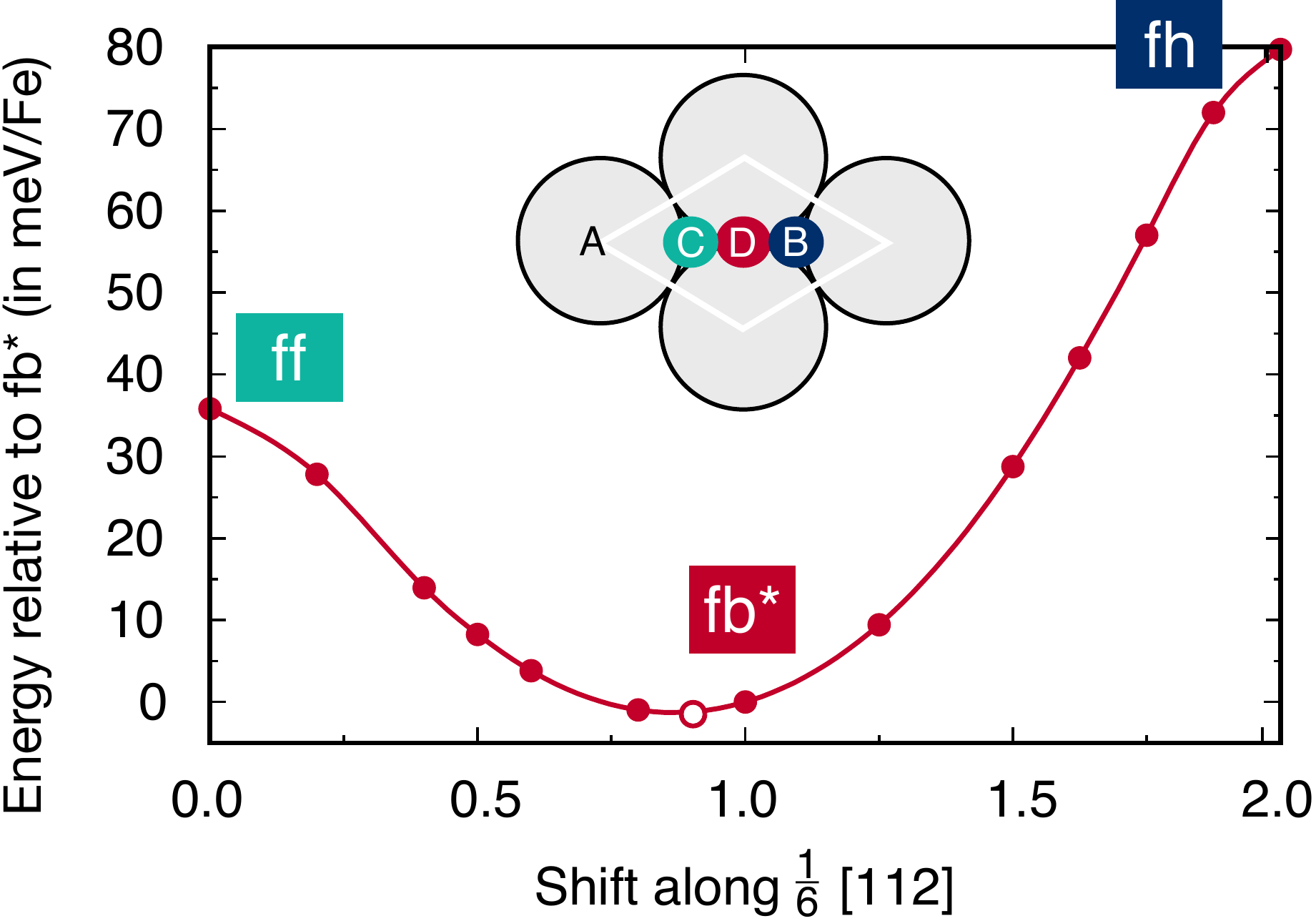}
\caption{(Color online) 
Total energies along the transformation paths from ff and fh to fb$^*$-stacking. The line is a guide to the eye.}
\label{fig:transformation}
\end{figure}

Therefore, starting from the ff-stacking and fh-stacking, we continuously shifted the Fe@vac-layer along the [112] direction relative to the fcc (111)-plane of the Fe@Ir-layer until we reached the lattice sites of the fb$^*$-stacking. In the inset of Fig.~\ref{fig:transformation}, these movements correspond to shifts from position C (ff-stacking) and position B (fh-stacking) to position D (fb$^*$-stacking), respectively. Thus, we follow the paths ff~$\rightarrow$~fb$^*$~$\leftarrow$~fh.

The corresponding energies can be found in Fig.~\ref{fig:transformation}. There are no energy barriers to overcome for reaching the fb$^*$ ground state, at least in the 0 K limit of our calculations. This means actually that both the ff and fh-stacking are mechanically unstable. Small distortions towards fb$^*$ shall lead to a phase transition. 

At higher temperatures this simple picture may change though. As was shown previously for the martensitic phase transition in bulk titanium,\cite{Petry1991} structures, which are mechanically unstable at 0 Kelvin, can be stabilized at elevated temperatures by phonon contributions. Such study is beyond the scope of this communication though.

Moreover, we observe that the fb$^*$-stacking ground state is actually degenerate. A small off-centering towards the direction of the ff-stacking by 0.08 \AA\ leads to a slightly lower energy by about 1.4 meV per Fe-atom as compared to fb$^*$. This result is confirmed by supercell calculations.\cite{Hauptmann2018} \\

We recently found that at specific growth conditions even more complex bcc-like superstructures resulting in zigzag patterns may be thermodynamically stable.\cite{Hauptmann2018}

\subsection{Magnetic interactions}

We investigate next the effect of the change of symmetry on the magnetic interactions for three  different stackings: ff, fb$^*$ and fh. In addition, we included the hf-stacking to check for the influence of the Fe@Ir-layer. We first present the energy dispersion curves of spin spirals without SOC, which are fitted to the Heisenberg model to obtain exchange constants within and between the two magnetic Fe-layers. We then show the effect of the double-layer stacking on the magnetocrystalline anisotropy. Finally, we investigate the Dzyaloshinskii-Moriya interaction in detail.

\subsubsection{Total magnetic exchange: Mapping of energy dispersions on extended Heisenberg model}

\begin{figure}[tb]
\centering
\includegraphics[width=1.0\columnwidth]{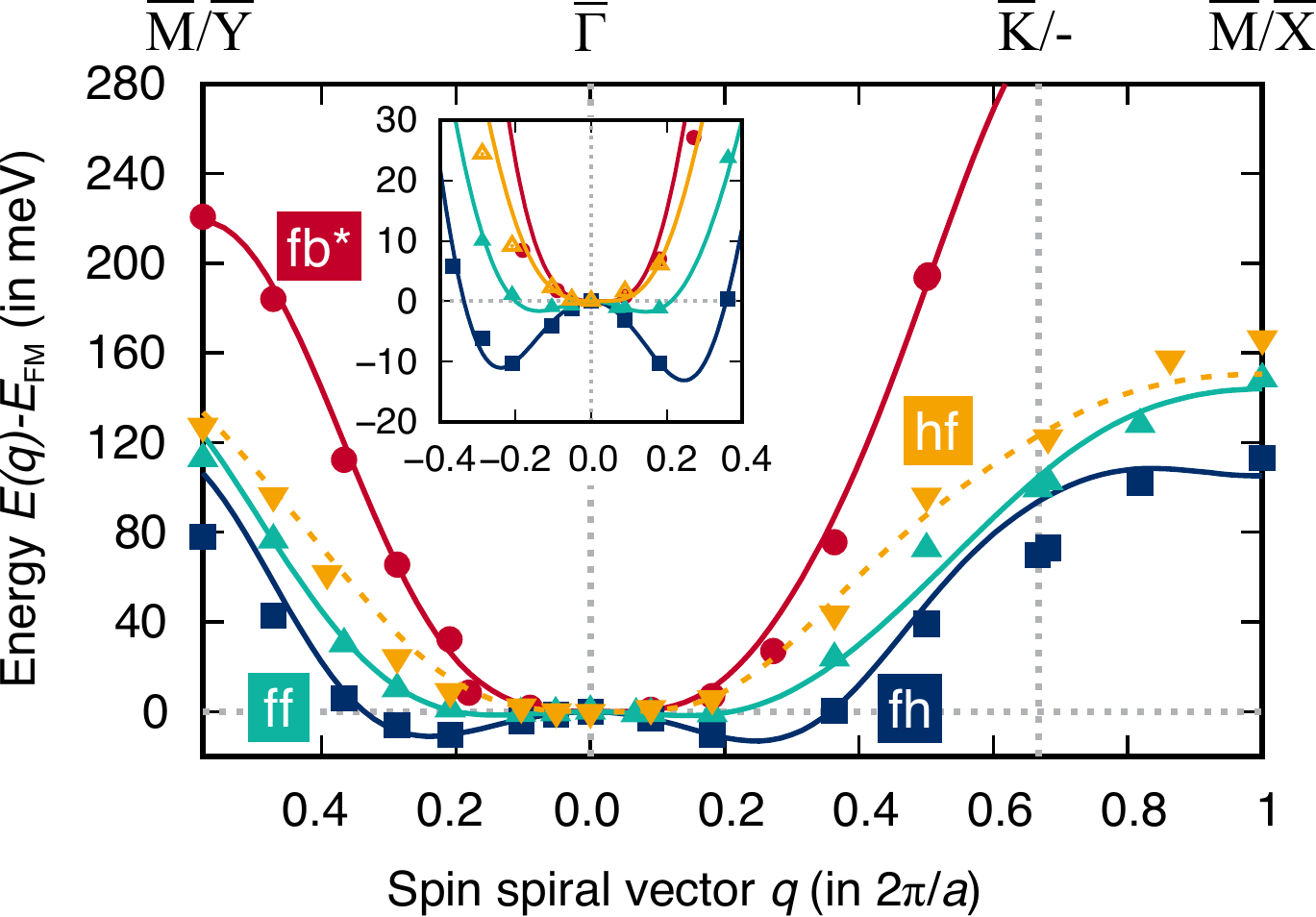}
\caption{(Color online) 
Spin spiral energy dispersions in the stackings ff, fb$^*$, hf and fh. Exchange stabilized spin spirals are found in ff and fh, while fb$^*$ and hf exhibit a ferromagnetic ground state.}
\label{fig:total-exchange}
\end{figure}

The total magnetic exchange interaction can be studied when the spin spiral is propagating within both Fe layers as well as the Ir substrate. Figure~\ref{fig:total-exchange} shows the energy dispersion curves of the four investigated Fe double-layer stackings. Two types of dispersion curves can be distinguished. In the case of ff (turquoise) and fh (dark blue), the magnetic exchange interactions favor spin spirals. The dispersion curve of fh has a deep minimum at $E=-13.1$~meV/Fe with $q=0.24\times (\tfrac{2\pi}{a})$ which corresponds to a spin spiral wavelength of $\lambda=1.1$~nm. Although the overall symmetry does not change, when the hcp stacking of Fe@vac is replaced by the fcc stacking, the energy minimum of the dispersion curve of ff is reduced to $E=-1.8$~meV/Fe and the wavelength increases to $\lambda=1.9$~nm with $q=0.14 \times (\tfrac{2\pi}{a})$. These two stackings have a spin spiral ground state.

On the other hand, the dispersion curves of hf (yellow) and fb$^*$ (red) exhibit an energy minimum at $q=0.0\times (\tfrac{2\pi}{a})$ which corresponds to the ferromagnetic state. In the bcc-like fb$^*$-stacking, the N\'{e}el state and the row-wise antiferromagnetic state at the BZ edges become extremely unfavorable ($E(\bar{\mathrm{X}})$=437~meV and $E(\bar{\mathrm{Y}})$=221~meV, which are more than twice as large as the corresponding energies for the close-packed stackings).\\

The occurence of isolated skyrmions depends on the rise of the dispersion curve close to $q=0.0\times (\tfrac{2\pi}{a})$ which can be estimated by using the effective nearest neighbor exchange constant $J_{\mathrm{eff}}$ as explained in section~\ref{Exch_mag}.\cite{Dupe2016} The values of $J_{\mathrm{eff}}$ (see Tab.~\ref{tab:J_eff}) confirm the qualitative findings of Fig.~\ref{fig:total-exchange}. They are in the same range as the values found in the Pd/Fe/Ir(111) ultra-thin films, where $J_{\mathrm{eff}}$ of $-2.3$ and $+4.4$~meV were reported depending on the stacking of the Pd overlayer.\cite{Dupe2016a} The stackings ff and hf possess dispersion curves with energy minima along each direction $\bar{\Gamma}$-$\bar{\mathrm{K}}$ and $\bar{\Gamma}$-$\bar{\mathrm{M}}$ which correspond to $J_\mathrm{eff}<0$. 
On the other hand, the hf and fb$^*$ stackings possess positive slopes and $J_\mathrm{eff}>0$. The disperion curve of hf rises even faster close to $\bar{\Gamma}$ than the one of fb$^*$, although the opposite is true further from $\bar{\Gamma}$. In both stackings, the magnetic exchange interaction favors a ferromagnetic ground state. 
In all stackings, $J_\mathrm{eff}$ are small, which facilitates the presence of isolated skyrmions at finite magnetic fields, as we could demonstrate previously for Pd/Fe/Ir(111).\cite{Dupe2014}

We want to analyze in more detail how the exchange-induced spin spirals in ff and fh develop by inspecting the exchange contributions from each Fe layer separately. \\

\begin{table}[tb]
\centering
\begin{ruledtabular}
\caption{\label{tab:J_eff}Effective next nearest neighbor exchange coefficients $J_\textrm{eff}$ corresponding to the dispersion curves of Fig.\ref{fig:total-exchange} for $|\bm{q}| < 0.1\times(\tfrac{2\pi}{a})$. All values are given in meV/Fe.}
\begin{tabular*}{\hsize}{ldddd}
double-layer stacking & \mathrm{ff} & \mathrm{fb^*} & \mathrm{fh} & \mathrm{hf}\\
 \hline
$J_\mathrm{eff}$ & -3.0 & +1.8 & -6.9 & +3.2 \\
\end{tabular*}
\end{ruledtabular}
\end{table}

\begin{figure}[tb]
\centering
\includegraphics[width=1.0\columnwidth]{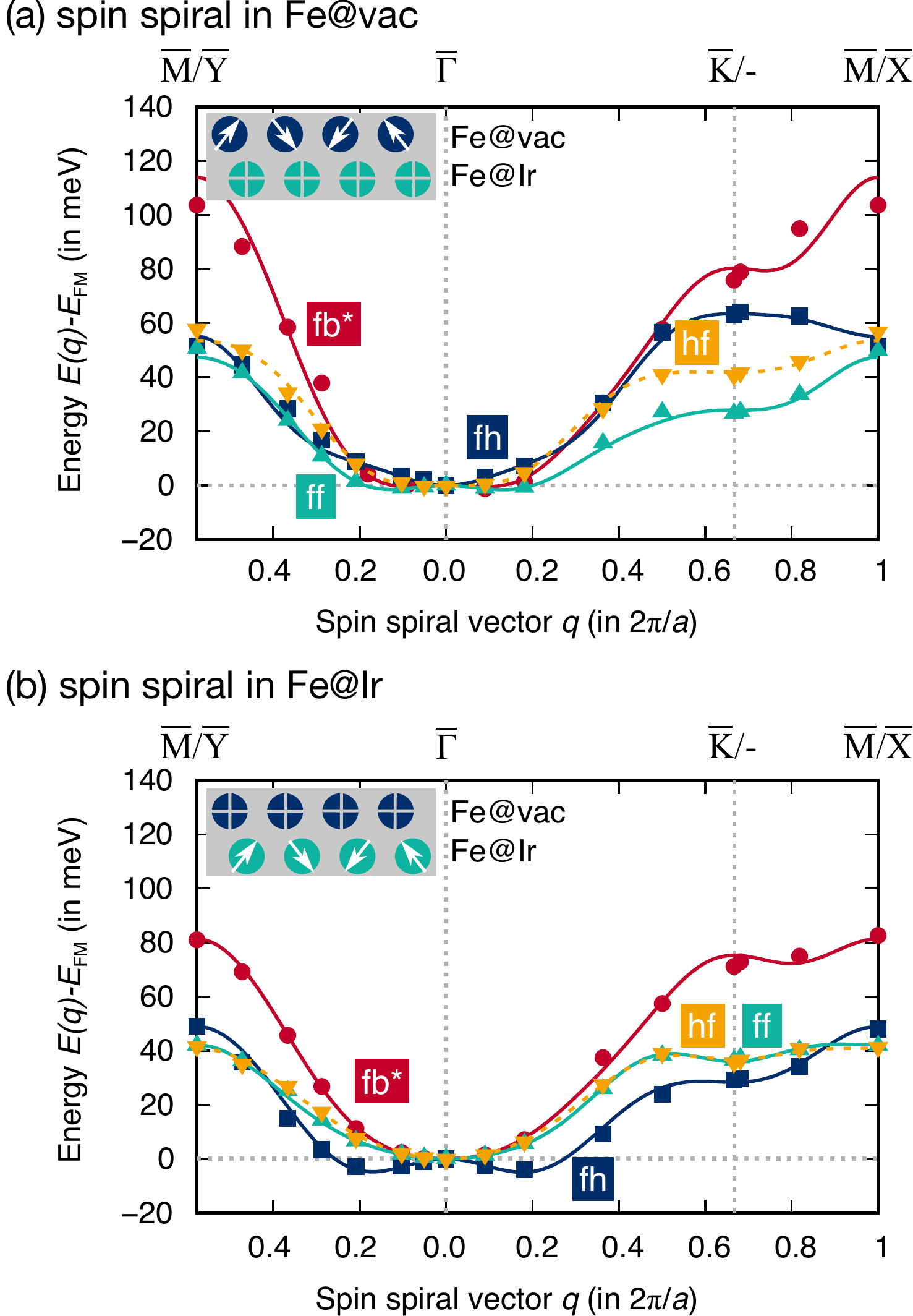}
\caption{(Color online) 
Layer-dependent spin spiral energy dispersions from DFT without SOC in the stackings ff, fb$^*$, hf and fh. In (a), the spin spiral propagates only in the Fe@vac-layer exposed to the vacuum, but not in the interfacial Fe@Ir-layer. In this spin configuration, the spin spiral is only stable in stacking ff, all other stackings favour the ferromagnetic state. In (b), the spin spiral propagates in the interfacial Fe@Ir-layer, but not in the Fe@vac-layer at the surface. In this case, we find a deep energy minimum for the spin spiral state in the hf-stacking, while the others remain ferromagnetic. }
\label{fig:layer-exchange}
\end{figure}

The layer dependent energy dispersion curves are shown in Fig.~\ref{fig:layer-exchange}, where a spin spiral was imposed only in one of the Fe-layers and the Ir-substrate, while spins in the other Fe-layer were kept parallel to each other and perpendicular to the rotational plane of the spin spiral resulting in a mixed spin configuration (spin spiral + ferromagnetic alignment). 

Figure~\ref{fig:layer-exchange}(a) shows the dispersion curves of the different stackings with a spin spiral in the layer of Fe@vac. All curves possess a flat energy dispersion close to the $\bar{\Gamma}$-point. This indicates that the magnetic exchange interaction in this layer is generally frustrated. Only the ff-stacking possesses a spin spiral ground state in this mixed spin configuration, while the remaining stackings are fully ferromagnetic.

Figure~\ref{fig:layer-exchange}(b) shows the dispersion curves for spin spirals in the layer of Fe@Ir. The dispersion curves of hf and ff are almost identical, although the local structure of the investigated layer is different. The curves of ff,fh and fb* rise faster close to the $\bar{\Gamma}$-point than in the Fe@vac case indicating less frustration. However, the dispersion curve of the fh stacking shows a deep minimum at $-5$~meV/Fe for $q=0.2\times (\tfrac{2\pi}{a})$.

Thus, the spin spirals in the ff and fh double-layers have different origins. In fh, its formation is driven by the Fe@Ir-layer, while in ff, it emerges from the Fe@vac-layer. 

It is interesting that the strong FM behavior of the fb$^*$ stacking observed in Fig.~\ref{fig:total-exchange} is lost for the single layer spin spirals (since $E(\bar{\mathrm{K}})=80$ and 110 meV/Fe) and that the 3-fold symmetry of the BZ is also recovered (since $E(\bar{\mathrm{X}})=E(\bar{\mathrm{Y}})$).\\

In order to analyze in detail all magnetic exchange interactions in the different stackings, we provide in Tab.~\ref{tab:J_ij} the fitted nearest neighbor interactions $J_{ij}$ up to the fifth and the fourth nearest neighbor in $J^{\parallel}$ and $J^{\perp}$, respectively.

All intra-layer exchange constants $J^{\parallel}$ show a high degree of frustration. They oscillate between ferromagnetic (FM) and antiferromagnetic (AFM) coupling depending on the shell number. In both Fe-layers, $J_1^{\parallel}$ is positive which shows that the FM state is more stable than the AFM states but the $J^{\parallel}$ beyond the first nearest neighbors can become negative.

All $J^{\parallel}$ have values below 12 meV/Fe which is between $J_1=5.7$ meV/Fe for Fe/Ir(111)~\cite{nphys2045} and $J_1$ is in the range of 13 to 14 meV/Fe for Pd/Fe/Ir(111). \cite{Dupe2014}   
Nevertheless, the inter-layer exchange constants $J^{\perp}$ can reach up to $J_1^{\perp}=62.7$ meV/Fe and $J_2^{\perp}=18.1$ meV/Fe in the case of fb$^*$. These values are much higher than $J_1^{\perp}=24.73$ meV/Fe for $[$Rh/Pd/2Fe/2Ir$]_1$. \cite{Dupe2016} This strong inter-layer coupling is the origin of the marked FM character of the fb$^*$ Fe double-layer observed in Fig.~\ref{fig:total-exchange}.  
These large values are originating from the symmetry lowering of the (110) surface where the first and the second shell contains only two atoms. They will induce a large anisotropy in the magnetic exchange interaction which can explain the occurrence of non-centrosymmetric skyrmions.\cite{Hsu2016,Hagemeister2016a,Hsu2017,Hauptmann2018} 

\begin{table}[tb]
\centering
\begin{ruledtabular}
\caption{\label{tab:J_ij}Heisenberg exchange coefficients $J_{ij}$ as obtained from fits to the spin Hamiltonians Eq.~\eqref{eq:Ham}. All values are given in meV/Fe.}
\begin{tabular*}{\hsize}{ldddd}
$J_i$ & \mathrm{ff} & \mathrm{fb^*} & \mathrm{fh} & \mathrm{hf}\\
 \hline
$J_1^{\parallel,\mathrm{Fe@vac}}$  & +5.0  & +12.3 &  +7.8 &  +6.1\\
$J_2^{\parallel,\mathrm{Fe@vac}}$  & +1.1  &   +1.3 &   -1.8 &  +1.4 \\
$J_3^{\parallel,\mathrm{Fe@vac}}$ &  -2.3  &    -5.2 &   -1.0 &   -1.1\\
$J_4^{\parallel,\mathrm{Fe@vac}}$  & +0.2  &   +0.9 &  +0.1 &   -0.2\\
$J_5^{\parallel,\mathrm{Fe@vac}}$ &  -0.5  &    -1.2 &  +0.7 &   -0.4\\
\hline
$J_1^{\parallel,\mathrm{Fe@Ir}}$  & +5.3  &  +9.6    &  +6.0 &  +4.9\\
$J_2^{\parallel,\mathrm{Fe@Ir}}$ & +0.1  &  +0.02  &  +0.5 &  +0.4\\
$J_3^{\parallel,\mathrm{Fe@Ir}}$  & -0.7  &  -2.3    &  -2.7 &   -0.4\\
$J_4^{\parallel,\mathrm{Fe@Ir}}$  & -0.3  &  +0.6    &  -0.1 &   -0.2\\
$J_5^{\parallel,\mathrm{Fe@Ir}}$ & +0.5  &  -0.6    &  -0.2 &  +0.2\\
\hline
$J_1^\perp$  & +24.0  &  +62.7 &  +12.1 & +22.1\\
$J_2^\perp$ &   +0.1  &  +18.1 &    +0.9 &   +2.0\\
$J_3^\perp$  &   -8.1  &    -6.1 &     -8.6 &   -5.8\\
$J_4^\perp$  &   +2.9  &    -2.3 &    +2.5 &  +1.3\\
\end{tabular*}
\end{ruledtabular}
\end{table}

To summarize, in ff- and fh-stackings, spin spirals are stabilized by magnetic exchange interaction. Their formation can be driven by only one of the Fe-layers. In contrast, in fb$^*$- and hf-stacking a flat dispersion curve is observed close to the $\bar{\Gamma}$-point indicating high magnetic frustration in these structures. In the latter structures, spin spiral ground states might be stabilized by the DMI, as we will see later. We will now consider the SOC contributions to the total energy.

\subsubsection{SOC contribution to collinear states: magnetocrystalline anisotropy}

The MAE is determined by calculating the SOC contribution to the total energy when all spins are pointing parallel along the $\mathbf{x}$, $\mathbf{y}$ and $\mathbf{z}$ directions. The MAE is then determined by the energy difference $E_i-E_{\mathrm{min}}$ with $i=x,y,z$. The easy axis or easy plane is the axis or plane along which the SOC contribution to the total energy is lowest, i.e. $E_{\mathrm{min}}$. 

\begin{figure}[tb]
\centering
\includegraphics[width=0.9\columnwidth]{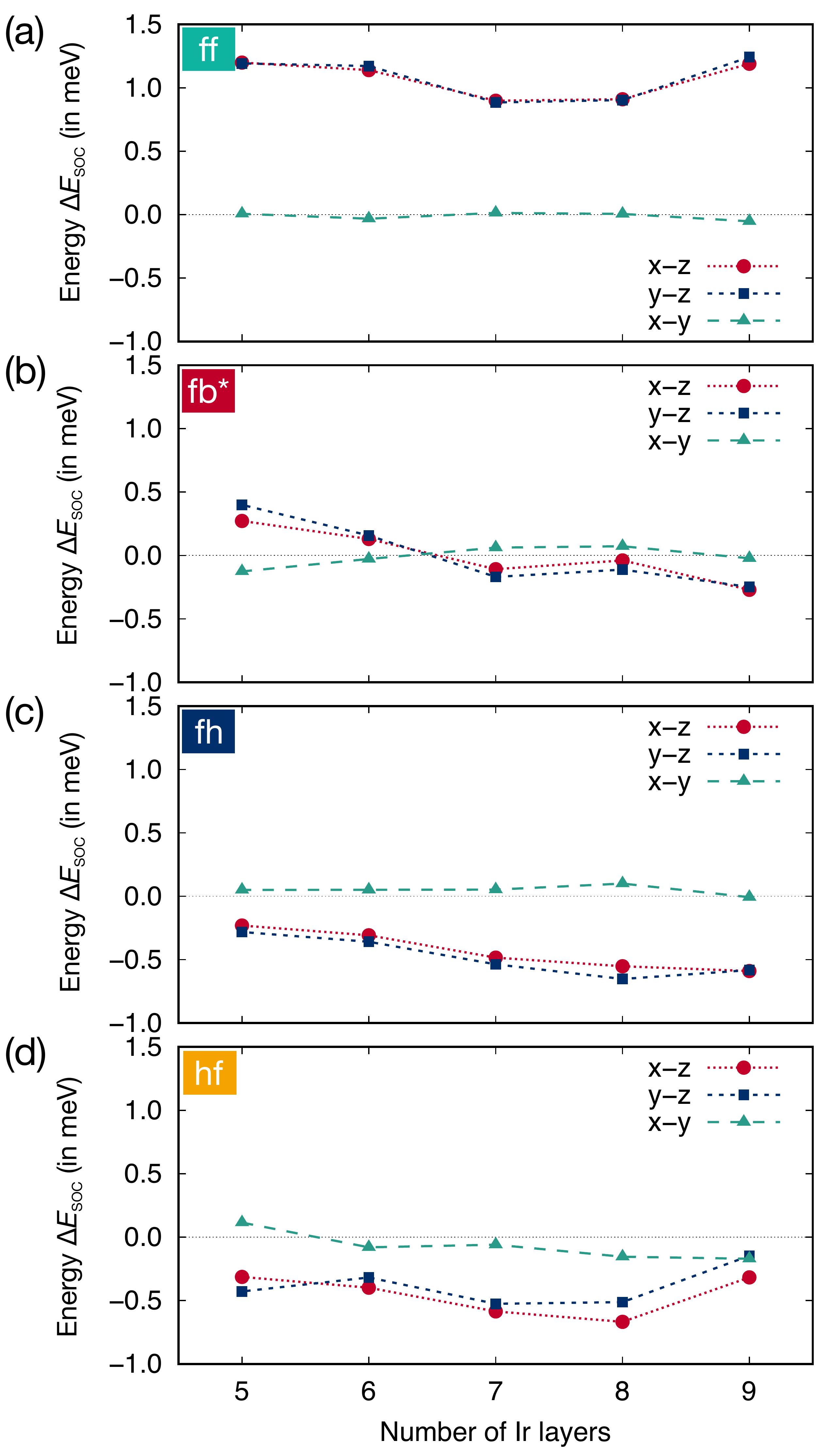}
\caption{Energy difference between FM states when SOC is applied along the $x$ and $z$ direction (dashed green triangle), along the $y$ and $z$ direction (dashed blue square) and along the $x$ and $z$ direction (dashed red dot) as a function of the the number of Ir layers in which SOC is applied in the calculation for the ff (a), fb$^*$ (b), fh (c) and hf (d).}
\label{fig:mae_by_layer}
\end{figure}

Since we perform calculations for asymmetric films there can be unphysical contributions from the Ir terminated side of the film. It is therefore important to explore the dependence on the MAE on the number of Ir MT spheres in which SOC is applied as shown in Fig. \ref{fig:mae_by_layer}. The MAE is rather independent from the number of Ir MT for the ff (a), fh (c) and hf (d) stackings, for which the energy difference oscillates around the values 1.2~meV, $-0.5$ meV and $-0.5$ meV, respectively. However, the MAE is much smaller for the fb$^*$ stacking and its variation oscillates around 0 meV. We approximate the MAE of fb$^*$ to 0.1~meV.

\begin{figure}[tb]
\centering
\includegraphics[width=0.9\columnwidth]{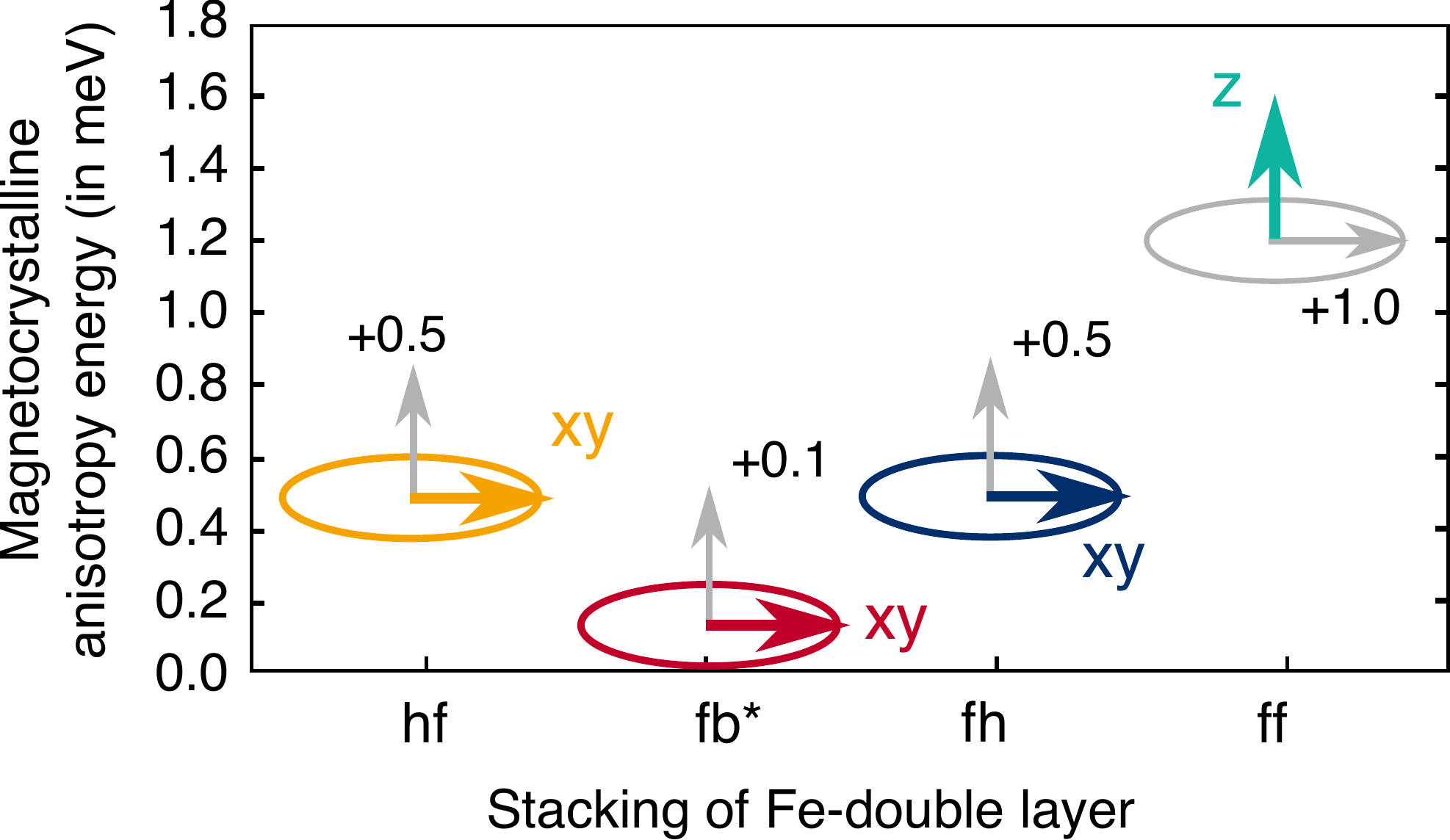}
\caption{(Color online) 
Shown are the preferred (in color and bold) and disfavoured spin orientations (in grey and fine) for the stackings hf, fb$^*$, fh and ff. 
The preferred orientation of spins changes from an easy-plane in hf,  fb$^*$ and fh-stacking to an out-of-plane easy-axis in the ff-stacking. 
The anisotropy coefficients $K_1$ increase at the same time from 0.1 meV in fb$^*$ to 1.2 meV in ff.}
\label{fig:mae}
\end{figure}

In Fig.~\ref{fig:mae} we summarize the MAE calculated for the four stackings ff, fb$^*$, fh and hf. The coloured axes and planes indicate the easy axis or easy plane, respectively. Numbers indicate the anisotropy coefficients $K_1$ with their associated directions. All energies are in the typical energy range of few meV for ultrathin-film systems. For ff we find an easy-axis pointing out-of-plane. Whereas fb$^*$, hf and fh possess an easy-plane in the basal plane of the film. The anisotropy coefficients $K_1$ in the easy-axis system is 1.0~meV (ff). The systems with the easy-plane have coefficients of $-0.1$ meV (fb$^*$) and $-0.5$ meV (fh and hf). 

It is surprising that the centered rectangular structure fb$^*$ posesses an easy-plane, where $x$- and $y$-directions are degenerate, although these directions are not equivalent by symmetry.

In the limit of small $\bm{q}$, the effect of the MAE on the stability of a spin spiral is a constant energy shift of the whole dispersion curve by $+\tfrac{1}{2}K_1$, i.e. it equally disfavors any kind of rotation of the magnetic moments.

\subsubsection{SOC contribution to non-collinear states: Dzyaloshinskii-Moriya interaction}

The DMI originates from SOC in non-centrosymmetric systems. 
In our systems, the broken inversion symmetry comes from the Ir-Fe interface, while the strong SOC originates from the Ir $5d$-states. 
The DMI is thus expected to be dominated by the atomic SOC contributions from the Ir(111) substrate.

\begin{figure}[tb]
\centering
\includegraphics[width=1.0\columnwidth]{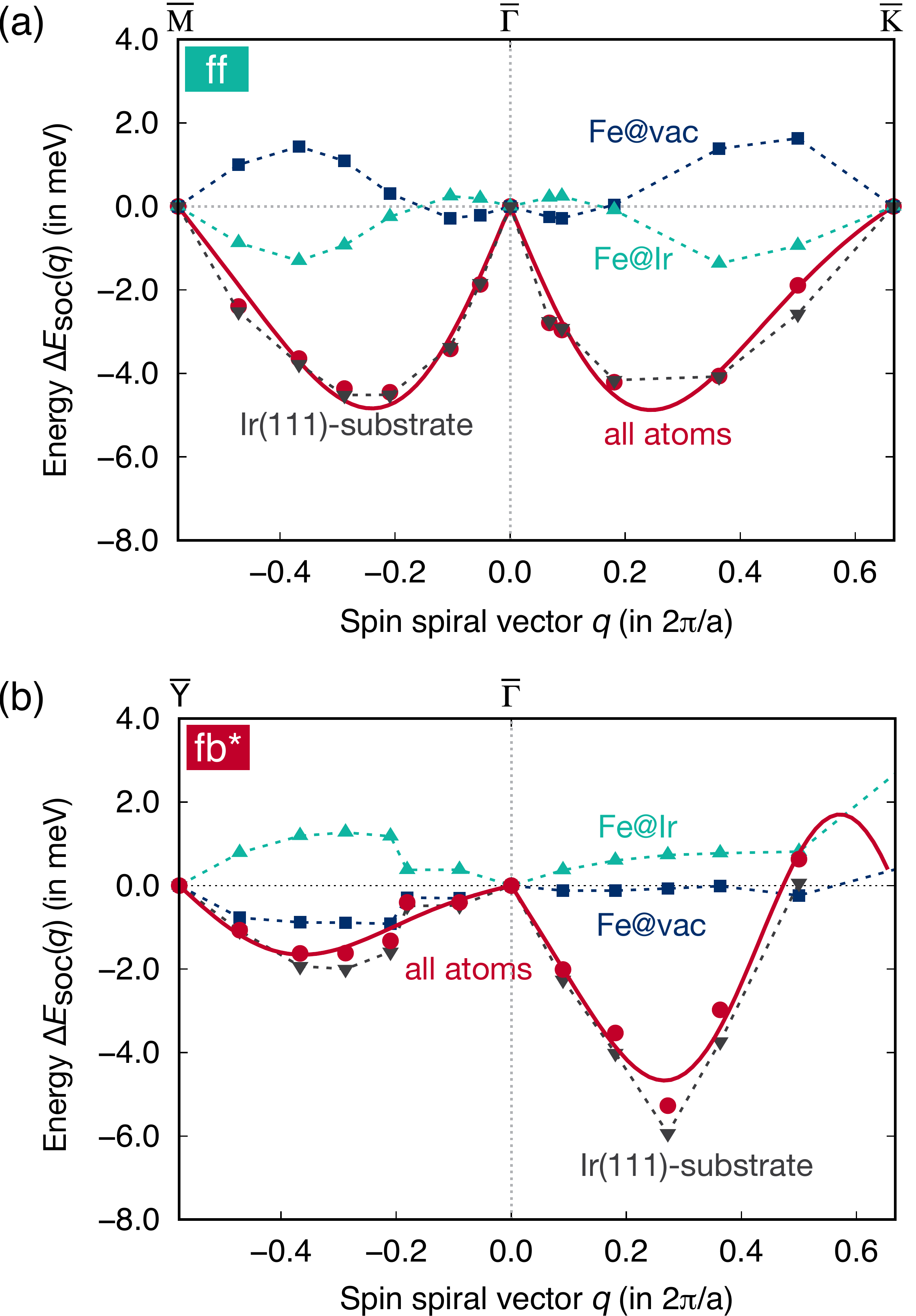}
\caption{(Color online) 
Spin-orbit coupling contribution to energy dispersion of cyloidal flat spin spirals related to the Dzyaloshinskii-Moriya interaction along high symmetry lines for stackings (a) ff and (b) fb$^*$. 
Given are the contributions of the different atoms as well as the fit for the total energy considering DMI up to the third nearest neighbor shell (see Tab.~\ref{tab:DMI_effective}). 
The DMI is considerably suppressed in the $\bar{\Gamma}$-$\bar{\mathrm{Y}}$ direction of the fb$^*$-stacking.}
\label{fig:DMI}
\end{figure}

Figure~\ref{fig:DMI} shows the total and atomic SOC contributions to the energy dispersions of spin spirals in the ff and fb$^*$ stackings. 
Figure~\ref{fig:DMI}(a) shows the atomic SOC contributions for the ff stacking. 
This stacking is also representative for the other 3-fold symmetric stackings (therefore, fh and hf are not shown). 
As expected, the SOC respects the 3-fold symmetry: Its total amplitude (red dots and solid lines) does not depend on the propagation direction of the spin spiral and is maximum for the 90$^{\circ}$ spin spirals at $q=-0.29\times (\tfrac{2\pi}{a})$ corresponding to $\mathbf{q}_{\mathrm{cart}}=(0,\tfrac{1}{2\sqrt{3}},0)$ and $q=+0.33\times (\tfrac{2\pi}{a})$ corresponding to $\mathbf{q}_{\mathrm{cart}}=(\tfrac{1}{3},0,0)$.
From the atomic SOC contributions we can see that the total SOC energy is dominated by contributions of the Ir substrate. 
The contributions of the two Fe layers (turquoise triangles and blue squares) are of opposite sign and therefore, cancel each other out.

The situation is different for the fb$^*$ stacking as shown in Fig.~\ref{fig:DMI}(b). 
The total SOC energy along the direction $\bar{\Gamma}-\bar{\mathrm{X}}$ has the same amplitude as in the ff case. 
However, the SOC energy along $\bar{\Gamma}-\bar{\mathrm{Y}}$ is reduced by a factor of three as compared to $\bar{\Gamma}-\bar{\mathrm{X}}$. 
This reduction of SOC interaction along the $\mathbf{k}_\mathrm{y}$-direction may be surprising. 
Especially, if one assumes that the symmetry of the DMI is determined solely by the symmetry of the interface.  
Fe@Ir occupies in the fb$^*$-stacking the same sites with respect to the Ir substrate as in the ff and fh stackings. 
Therefore, only the surface Fe-layer (Fe@vac) reduces the symmetry of the ultrathin film, which is usually not expected to modify the hybridization of the interfacial Ir atoms.

In Tab.~\ref{tab:DMI_effective} we compare the DMI coefficients $D_\mathrm{eff}$ and $D_{1-3}$ acting on the interfacial Fe (Fe@Ir) of the stackings ff, fh, fb$^*$ and hf, which we obtained from fitting the total SOC energies.
For stackings ff and fh, we find $D_\mathrm{eff}$ of about 2.0~meV, whereas fb$^*$ and hf give about 1.2~meV. 
As $D_\mathrm{eff}$ in all stackings adopts positive values it favours clockwise-rotating spin spirals. 
These DMI coefficients are in the same range as those for the Fe monolayer on Ir(111)\cite{Heinze2011} and those of the Fe/Ir(111)-system with Pd-overlayers.\cite{Dupe2014}
The centered rectangular symmetry of the fb$^*$ stacking is reflected in the existence of a second $D_\mathrm{eff}$. The DMI in the Cartesian y-direction is strongly suppressed to $D_\mathrm{eff}=0.2$~meV, only one sixth of the value along the Cartesian $x$-direction. 

\begin{table}[t]
\centering
\begin{ruledtabular}
\caption{\label{tab:DMI_effective}DMI coefficients of Fe@Ir in all stackings determined along $\bar{\Gamma}$-$\bar{\mathrm{K}}$ and $\bar{\Gamma}$-$\bar{\mathrm{M}}$ direction for ff, fh, and hf and along $\bar{\Gamma}$-$\bar{\mathrm{X}}$ direction for fb$^*$ (along $\bar{\Gamma}$-$\bar{\mathrm{Y}}$ in parentheses). 
All values are given in meV.}
\begin{tabular*}{\hsize}{lcccc}
double-layer stacking & ff & fb$^*$ & fh & hf\\
 \hline
$D_\mathrm{eff}$ &  2.0 & 1.2 (0.2) & 2.0 & 1.3 \\
\hline
$D_1$  & 1.25  &  1.17 & 1.25 & 1.21 \\
$D_2$  & 0.16  &  0.90 & 0.16 & -0.05 \\
$D_3$  & 0.11  & -0.78 & 0.11 & 0.10 \\
\end{tabular*}
\end{ruledtabular}
\end{table}

In Tab.~\ref{tab:DMI_effective}, also the coefficients for the DMI up to the third next nearest neighbor are provided. The values of $D_1$ are very similar for all stackings. For the ff and the fh stackings, $D_{1-3}$ are quasi identical and they are all positive, which creates a $D_\mathrm{eff}$ of around 2 meV/Fe. The $D_\mathrm{eff}$ is significantly reduced to 1.2 meV/Fe for the hf and the fb$^*$ stackings due to the negative contribution of the $D_2$ and the $D_3$, respectively.
The $D_\mathrm{eff}$ of fh and ff differ due to small energy differences of the SOC contributions close to $\bar{\Gamma}$.\\  

\begin{figure*}[tb]
\centering
\includegraphics[width=2.0\columnwidth]{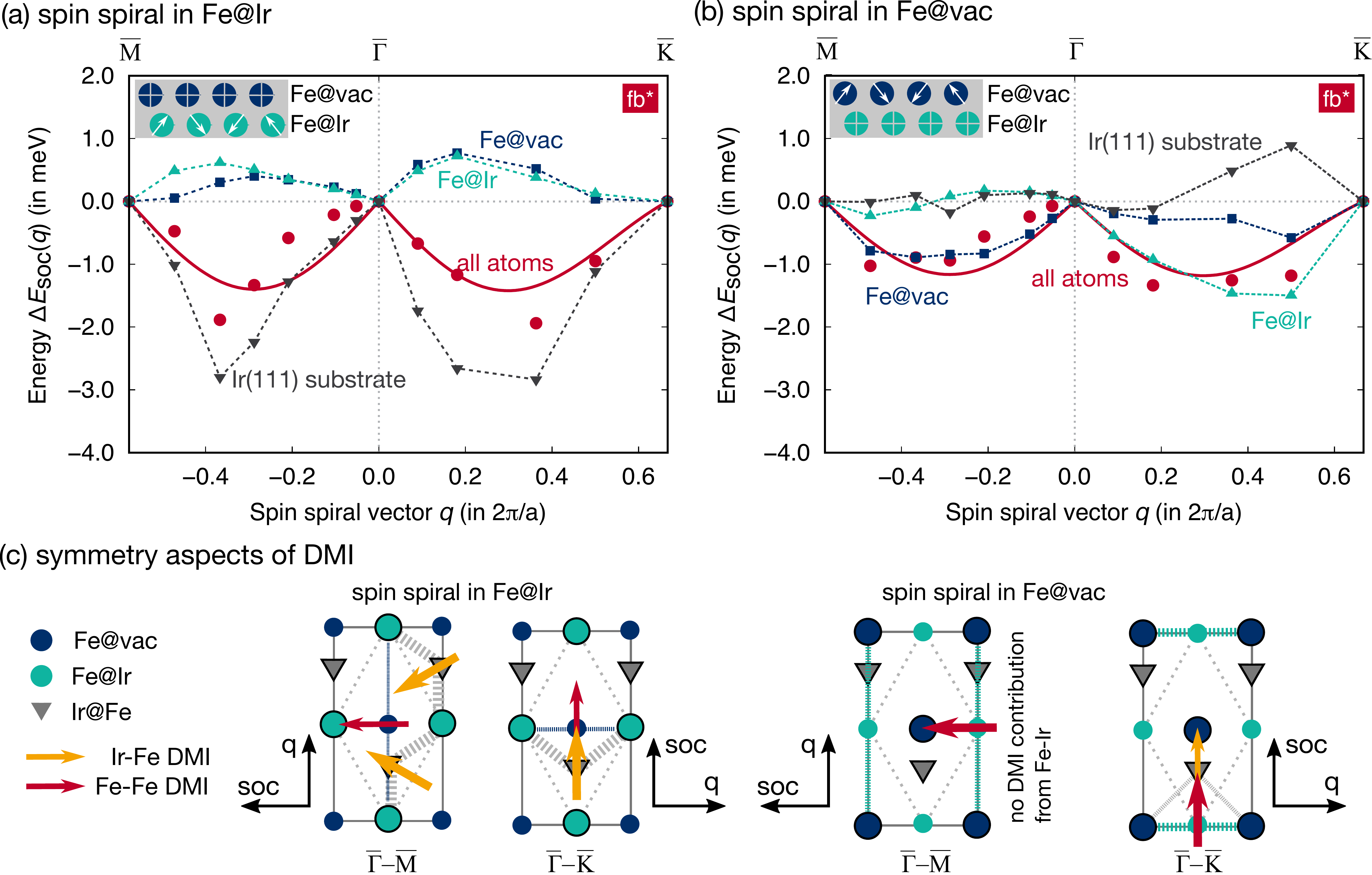}
\caption{(Color online) 
Total energy contributions to spin spiral energy in the fb$^*$ stacking dispersions due to SOC resulting from spin spirals in only one of the Fe-layers. Given are the atomic contributions and the total SOC energies. In (a), the spin spiral propagates in the interfacial Fe@Ir-layer, but not in the Fe@vac-layer at the surface. The magnitudes along both propagation directions are the same. Thus, the SOC possesses 3-fold symmetry in this spin configuration. The corresponding DMI coefficient is $D_1(\mathrm{Fe@Ir})=0.40$ meV. In (b), the spin spiral propagates in the Fe@vac-layer of the surface. The magnitudes along the two propagation directions are completely different. The SOC reflects the reduced symmetry of the bcc-like stacking. The obtained DMI coefficient is $D_1(\mathrm{Fe@vac})=0.34$ meV. In (c), we show how the Ir-Fe DMI  (yellow arrows) and Fe-Fe DMI (red arrows) arise from different 3-atom scattering events possible in the two different spin spiral configurations. Shown are the centered rectangular unit cells (grey solid line) in top view, along with the rhombic unit cell (dotted line) for orientation. The quantization axis of the SOC is always perpendicular to the spin spiral propagation direction which leads to cycloidals spin spirals along $\boldmath{q}$. The scattering partners are connected via dashed lines, in grey for Ir-Fe and in turquoise/dark blue for Fe-Fe interactions. Only for the spin spiral in Fe@vac along the $\bar{\Gamma}-\bar{\mathrm{M}}$ direction, there is no Ir-Fe DMI possible, which gives rise to the strong asymmetry in (b).}
\label{fig:DMI-decomposed}
\end{figure*}

In order to confirm that the symmetry of the DMI in the fb$^*$ stacking depends on the adsorption site of the surface Fe-layer (Fe@vac), we calculated the layer dependent SOC for two different spin spiral configurations as in Ref.~\cite{Dupe2016}. 
In the first configuration, the spin spiral propagates only in the Fe@Ir-layer and the Ir-substrate, but not in the Fe@vac-layer, where the magnetic moments are oriented parallel to each other (FM) and perpendicular to the rotation plane of the spin spiral. Here we calculate the SOC contribution from the Ir substrate on Fe@Ir. 
In the second configuration, the spin spiral propagates exclusively in the Fe@vac-layer and in the Ir-substrate, but not in the Fe@Ir-layer to obtain the SOC contribution from the Ir substrate on Fe@vac.
The results are presented in Fig.~\ref{fig:DMI-decomposed}. 
As each Fe layer taken isolated has 3-fold symmetry, we chose the trigonal reference for the Brillouin zone spanned by $\bar{\mathrm{M}}$, $\bar{\Gamma}$ and $\bar{\mathrm{K}}$. 
Fig.~\ref{fig:DMI-decomposed}(a) corresponds to the case where the spin spiral propagates in the Fe@Ir layer and Ir substrate, but not in the Fe@vac layer. 
It is interesting that indeed a 3-fold symmetric SOC is retained, since both $\bar{\Gamma}-\bar{\mathrm{M}}$ and $\bar{\Gamma}-\bar{\mathrm{K}}$ directions exhibit the same magnitude for both the total SOC and the individual atomic contributions. 
The $\Delta E_\textrm{SOC}(q)$ is dominated by the atomic contribution from the Ir substrate, whereas the atomic contributions from the two Fe layers are equally strong and of opposite sign as compared to the Ir substrate, giving an overall DMI coefficient of $D_1=0.40$~meV for this spin spiral configuration (red solid line).

However, when the spin spiral propagates in the Fe@vac layer and the Ir substrate [Fig.~\ref{fig:DMI-decomposed}(b)], the symmetry is reduced to c$m$, as was shown in Fig.~\ref{fig_symmetries2}(b). 
Therefore, we can expect the SOC to have a different amplitude depending on the direction of the propagation vector. 
Indeed, when the SOC is computed along the $\bar{\Gamma}-\bar{\mathrm{M}}$ direction, the contributions of Fe@Ir and the Ir-substrate are reduced to zero (turquoise and grey triangles, respectively) while the main contribution arises from Fe@vac  (compare the blue squares and the red dots).
Along the perpendicular direction $\bar{\Gamma}-\bar{\mathrm{K}}$, all atoms are contributing to the SOC, the dominating contribution originates from Fe@Ir, whereas the contributions of Fe@vac and the Ir-substrate counterbalance each other. Overall, the DMI is dominated by Fe-Fe interactions in this spin spiral configuration giving an overall DMI coefficent of $D_1=0.34$~meV.

We next analyze how  these differences between the two spin spiral configurations can be understood. 
In the model of Fert and Levy,\cite{Fert1980} the DMI is associated with a scattering process among three atoms, arranged in an isosceles triangle, where the scattering atom at the apex can be non-magnetic. This atomic configuration leads to cycloidal spin spirals which propagate in the plane of the ultra-thin film. 
Therefore, the available scattering partners for the two spin spiral configurations and propagation directions can be directly identified and are presented in Fig.~\ref{fig:DMI-decomposed}(c). 
The DMI associated with the interactions between the Ir substrate and Fe@Ir or Fe@vac and between Fe@Ir and Fe@va are indicated by yellow and red arrows, respectively, as well as the inter-layer connection lines between the three scattering atoms (dashed lines).
For orientation, besides the centered rectangular unit cell (grey solid line) also the rhombic unit cell is indicated (dotted line). 
In the first spin spiral configuration exhibiting the full 3-fold symmetry, shown on the left, we have strong Ir-Fe DMI and much weaker Fe-Fe DMI in both propagation directions of the spin spiral. 
Along $\bar{\Gamma}-\bar{\mathrm{K}}$ both DMI vectors are parallel to the quantization axis and contribute fully to the DMI.
Whereas in the direction $\bar{\Gamma}-\bar{\mathrm{M}}$, the two Ir-Fe DMI vectors are rotated by 30$^{\circ}$ with respect to the quantization axis.
Therefore, they both contribute only 50\% to the DMI, resulting overall in the same magnitude for the Ir-Fe DMI as in the direction $\bar{\Gamma}-\bar{\mathrm{K}}$.

The situation is different in the second spin spiral configuration shown in Fig.~\ref{fig:DMI-decomposed}(d). In the direction $\bar{\Gamma}-\bar{\mathrm{M}}$, no DMI contribution from Ir-Fe can be found, as no isoceles triangles formed by Fe@vac $-$ Ir@Fe $-$ Fe@vac exist. 
Therefore, in this direction only the weak Fe-Fe DMI is present.
In contrast to the $\bar{\Gamma}-\bar{\mathrm{K}}$ direction, where DMI from both interactions Ir-Fe and Fe-Fe can be found. 

To summarize, the absence of the Ir-Fe DMI in the Fe@vac spin-spiral due to the reduced symmetry is the origin of the reduced symmetry of the total DMI in the fb$^*$ stacking. Thus, the symmetry of the surface Fe-layer indeed determines the symmetry of the DMI, hence, not only the symmetry of the interface matters. This confirms that the DMI is anisotropic and could allow for the presence of antiskyrmions in this specific stacking.\cite{Hoffmann2017}\\

\subsection{Electronic Structure}

\begin{figure*}[tb]
\centering
\includegraphics[width=2.0\columnwidth]{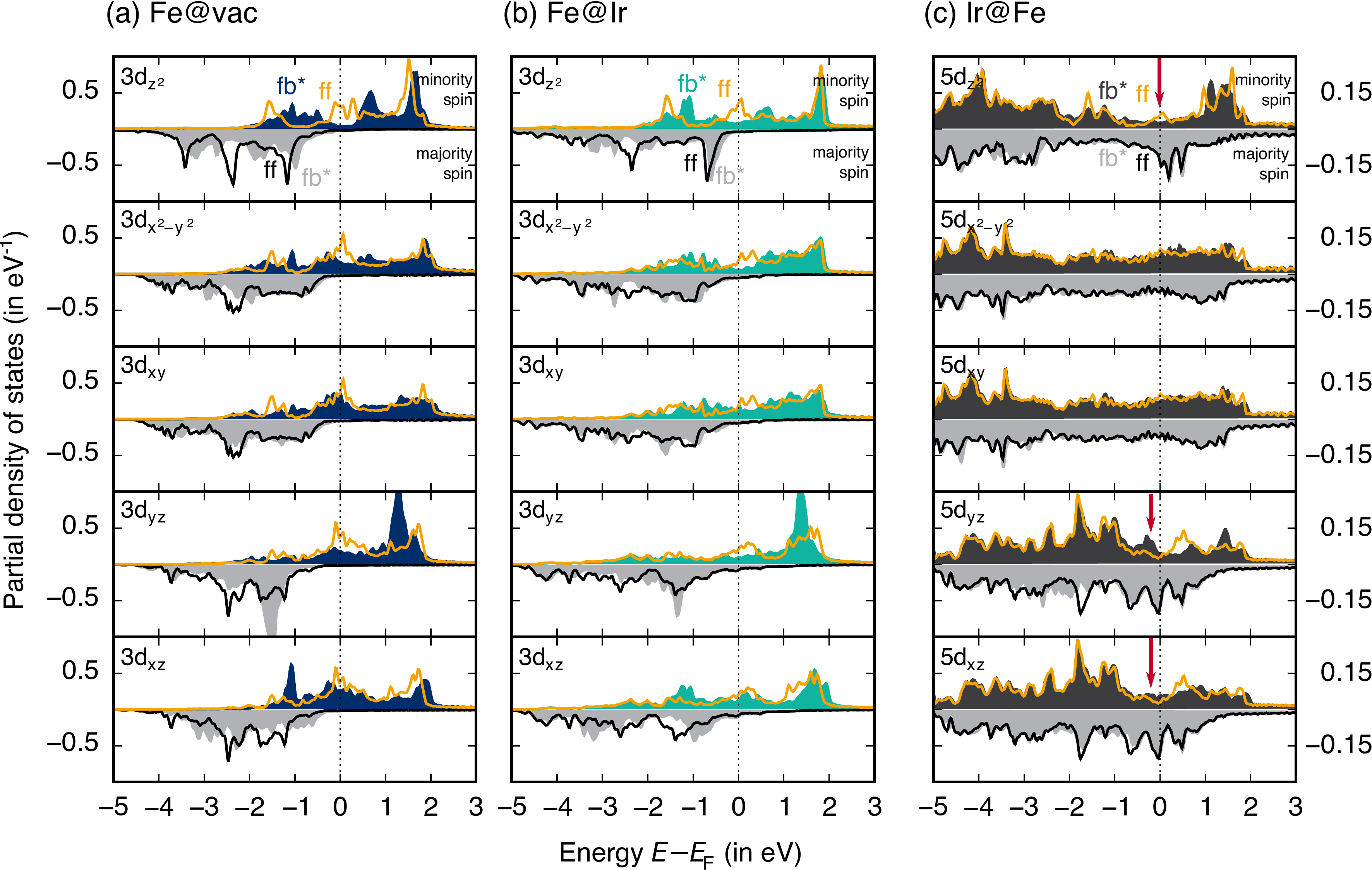}
\caption{(Color online) 
Orbital and angular momentum-resolved partial densities of states (PDOS) for the $3d$- tates of (a) Fe@vac, (b) Fe@Ir and for the $5d$ states of (c) Ir@Fe in the stackings ff (lines) and fb$^*$ (filled). 
Both majority spins (positive PDOS) and minority spins (negative PDOS) are given. The stacking of the Fe@vac layer changes the hybridization in the Ir@Fe layer as indicated by the arrows. 
Please note the different PDOS scales for Fe and Ir atoms.}
\label{fig:DOS}
\end{figure*}

In order to support the latter statements we present the orbital-resolved partial densities of states (PDOS) of the atoms Fe@vac, Fe@Ir and Ir@Fe in the stackings ff and fb$^*$. The PDOS are given in Fig.~\ref{fig:DOS}. 
Presented are the $3d$ orbitals of the Fe atoms and the $5d$ orbitals of the Ir atom. 
In each panel both the majority (greyscale, negative values) and minority spins (color, positive values) are given. 
We directly compare the PDOS of fb$^*$ (filled) and ff (lines) for each atom, orbital and spin channel.

The main changes arising from the structural differences between ff and fb$^*$-stackings occur in the PDOS of the Fe atoms in both spin channels. 
The majority spin states of the two Fe atoms are fully occupied. 
We observe the largest differences between the two stackings close to the Fermi energy which is populated by minority spin states. 
In ff, the orbitals $3d_\mathrm{x^2-y^2}$ and $d_\mathrm{xy}$ as well as $3d_\mathrm{yz}$ and $3d_\mathrm{xz}$ of both Fe@vac and Fe@Ir are degenerate. 
Moreover, all Fe $3d$ orbitals possess a peak at the Fermi energy. 
We have overall a three-peak pattern with a small peak at $-1.5$~eV (bonding states), and two large peaks around $E_\textrm{F}$ (non-bonding states) and $+1.5$~eV (anti-bonding states). \\
In fb$^*$, the degeneracy is lifted and the peak structure changes. 
We find two very broad peaks below and above the Fermi energy. Around $E_\textrm{F}$, the DOS adopts only small values, especially apparent in the $3d_\mathrm{z^2}$ and $3d_\mathrm{x^2-y^2}$ states. 
Thus, the states with non-bonding character vanish, which explains the enhanced thermodynamic stability of the fb$^*$-stacking as compared to the close-packed structures. 
Moreover, the degeneracy of the $3d_\mathrm{yz}$ and $3d_\mathrm{xz}$ states is lifted in a particular way. 
While the new $3d_\mathrm{xz}$ orbital rather has a three-peak structure with a large peak at the bonding states at $-1$~eV, the $3d_\mathrm{yz}$ orbital becomes mainly anti-bonding with a large peak at $+1.3$~eV. 
This is a result of the symmetry breaking in the centered rectangular unit cell of the fb$^*$ structure.
The hybridization between the two Fe layers is enhanced especially in the x- and z-direction (i.e. $3d_\mathrm{xz}$ and $3d_\mathrm{z^2}$ orbitals overlap strongly), where the interlayer atomic distances are minimal. 
Moreover, it is reduced in the $3d_\mathrm{yz}$ orbital, where the DOS is largest above the Fermi energy and which corresponds to an orientation, where interlayer atomic distances are maximal.
Thus, the changes in the electronic structures reflect the symmetry differences between the stackings and can also explain the enhanced thermodynamic stability of the fb$^*$-stacking.\\

Let us focus next on the electronic structure of the Ir-atoms in order to find indications for the reduction of DMI in the fb$^*$-stacking.
First, we find no effect of the Fe double-layer structure on the $5d_\mathrm{x^2-y^2}$ and $5d_\mathrm{xy}$ states as they have no out-of-plane component and are solely directed towards Ir-atoms of the same layer. 
This is in contrast to the remaining three orbitals, which are either uniquely directed towards the Fe atoms like $5d_\mathrm{z^2}$ or have at least out-of-plane contributions like $5d_\mathrm{yz}$ and $5d_\mathrm{xz}$. 
Both groups of orbitals are affected differently by the structural changes in the Fe@vac layer. 
In ff, we find a small peak at the Fermi energy in $5d_\mathrm{z^2}$ orbital, whereas fb$^*$ shows rather a minimum. 
For the $5d_\mathrm{yz}$ and $5d_\mathrm{xz}$ states the effect is reversed, but to a different degree.
We observe a peak three times larger than the PDOS of the ff-stacking at $-0.3$~eV for the $5d_\mathrm{yz}$ orbital, i.e. a strong increase of the PDOS just below $E_\textrm{F}$ in fb$^*$.
Whereas for the $5d_\mathrm{xz}$ orbital, the PDOS is doubled in this energy range as compared to the ff-stacking. 
Thus, the electronic states of the atoms in the Ir@Fe-layer "sense" the symmetry breaking due to the Fe@vac-layer and react with an unexpected lifting of degeneracy in the $5d_\mathrm{yz}$ and $5d_\mathrm{xz}$ orbitals. 

\subsection{Magnetic ground state}

\begin{figure*}[tb]
\centering
\includegraphics[width=2.0\columnwidth]{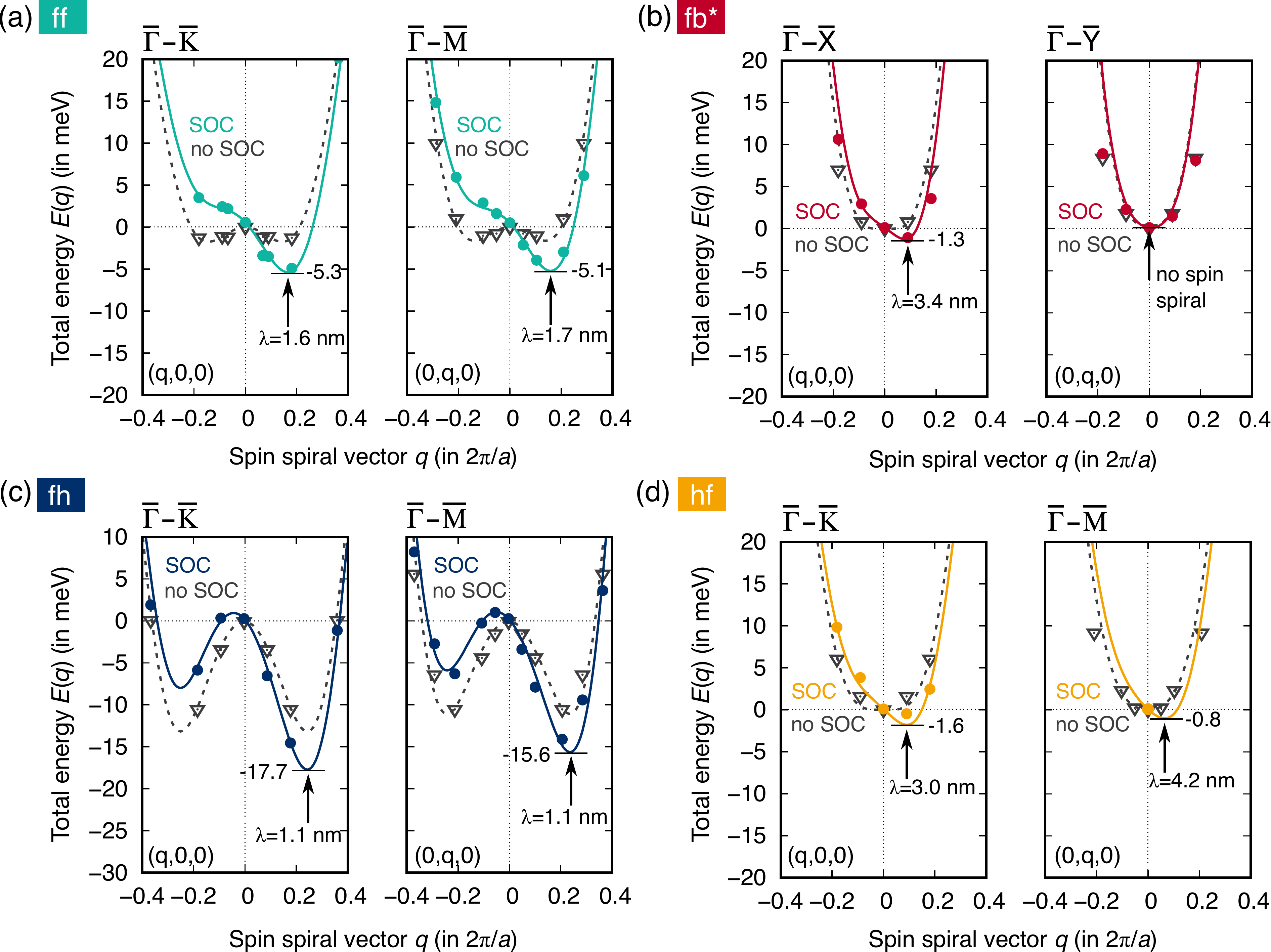}
\caption{(Color online) 
Full energy dispersions including SOC of flat cycloidal spin spirals from DFT relative to the FM state of stackings (a) ff, (b) fb$^*$, (c) fh and (d) hf along high symmetry directions $\bar{\Gamma}-\bar{\mathrm{K}}$ and $\bar{\Gamma}-\bar{\mathrm{M}}$. 
Given are values with (filled circles, colored solid lines) and without SOC  (open triangles, grey dashed lines). Positive (negative) values of $q$ indicate clockwise (anti-clockwise) rotating spin spirals.
In all stackings clockwise-rotating spin spirals are favoured. The contribution of the magnetocrystalline anisotropy to spin spirals amounts to $\tfrac{1}{2}K_1$. }
\label{fig:SpinSpirals}
\end{figure*}

Finally, we have determined all quantities in order to study the thermodynamic stability of spin spirals in the different structures at $T=0$~K.
In Fig.~\ref{fig:SpinSpirals} we present the energy dispersions with and without the contributions arising from SOC in positive and negative propagation directions along the high-symmetry lines of the BZ. 

As the DMI coefficients are positive, in all stackings clockwise-rotating spin spirals (along positive propagation directions) are favored. The contribution of the magnetocrystalline anisotropy amounts to $\tfrac{1}{2}K_1$ and leads to a constant energy shift of spin spirals with respect to the FM state ($q=0\times (\tfrac{2\pi}{a})$).
Indicated are the energies of the spin spirals relative to the ferromagnetic states and their wave lengths. 
All stackings possess spin spiral ground states. 

The most isotropic structure is the ff-stacking, given in \ref{fig:SpinSpirals}(a). Here, both energies (about $-5.8$~meV) and wavelength (1.7 nm) of the spin spirals along the Cartesian $\overline {\Gamma \mathrm{K}}$ and $\overline {\Gamma \mathrm{M}}$ directions are quasi identical. Thus, in this structure no preferred propagation direction exists. As the spin spirals are already stabilized by the exchange interaction, the additional DMI gives rise to the asymmetry of the dispersion curve with respect to the chirality (clockwise or anti-clockwise). Here the DMI favors clockwise-rotating cycloidal spin spirals. Moreover, the DMI shifts the energy minimum to larger $q$-vectors, thus accelerating the spin spiral.

Completely different is the result of the fb$^*$-stacking, presented in \ref{fig:SpinSpirals}(b). Here, the anisotropy of the DMI is also reflected in a strong anisotropy of the magnetic texture. Only one spin spiral can exist in the fb$^*$ structure, which propagates along the cartesian $\overline {\Gamma \mathrm{K}}$ direction. It has a relatively long wave length of 3.5 nm and a shallow energy minimum at $-1.2$ meV.  The DMI in the $\overline {\Gamma \mathrm{M}}$ direction is too small to stabilize a spin spiral along this direction. Therefore this stacking is expected to exhibit a spin spiral only in three directions in space.

Very short wavelength spin spirals with low energy result in the fh-stacking shown in \ref{fig:SpinSpirals}(c). Spin spirals along the $\overline {\Gamma \mathrm{K}}$ and $\overline {\Gamma \mathrm{M}}$ direction should both have a wavelength of 1.1 nm, but the spin spiral along the $\overline {\Gamma \mathrm{M}}$ direction is energetically favoured by almost 2 meV. Also in the hf-stacking Fig.\ref{fig:SpinSpirals}(d), we find a preference for a spin spiral propagating along the $\overline {\Gamma \mathrm{K}}$ direction. It has an energy of $-1.6$ meV and a wavelength of 3.0 nm as compared to the spin spiral along y-direction, which gives $-0.8$ meV and a much longer wavelength of 4.2 nm. 
The DMI stabilized spin spirals in fb$^*$ and hf have considerably longer wavelengths than the exchange stabilized spin spirals in ff and fh. 

Thus, all stackings favor spin spiral ground states with different symmetries. They are either stabilized by both magnetic exchange and DMI, as in ff and fh, or only by DMI, as in fb$^*$ and hf. There is a more or less strong preference for spin spirals along the  $\overline {\Gamma \mathrm{K}}$ direction depending on the stacking. In fb$^*$, along the $y$-axis no spin spiral can be stabilized at all, while in ff, spin spirals along $\overline {\Gamma \mathrm{K}}$ and $\overline {\Gamma \mathrm{M}}$ direction are degenerate.

Let us compare our results with previous STM experiments. In the supposedly ff-stacked regions a spin spiral of 1.2 nm was identified without a preferred propagation direction.\cite{Hsu2016} We also find for this stacking, that the spin spirals along the different propagation directions are degenerate, but the wavelength is 1.6 nm, which corresponds well with the experimental value. In the fb$^*$-stacked regions, spin spirals propagating only along the [100]-direction of the bcc-unit cell with a wavelength of 1.9 nm were reported. We can confirm this finding. In the fb$^*$-stacking, spin spirals are only found in the $\overline {\Gamma \mathrm{X}}$ direction, which corresponds to the [100] crystallographic axis of the bcc-unit cell. However, our wavelength of 3.4 nm is substantially larger than the experimental value. This is probably due to the differences in strain exposed on the Fe@vac-layer at the surface. We assumed a pseudomorphically strained bcc-like stacking, while SP-STM experiments reveal a complex surface reconstruction, which probably occurs in order to compensate for the large strain. We recently addressed this issue in another publication.\cite{Hauptmann2018}

\section{Conclusion}

We have investigated the interplay between the Fe double-layer stacking and the stabiliy and the magnetic interactions for 2Fe/Ir (111) ultra-thin films using density functional theory and mapping on an atomistic spin model.

We considered in total six different double-layer stackings: the hexagonal close-packed variants ff, fh, hf and hh, and two structures with a bcc (110)-like surface Fe-layer fb$^*$ and hb$^*$. 
We find that the fb$^*$ stacking with a bcc (110)-like surface layer is the most stable pseudomorphic structure, whereas the hexagonal close-packed stackings ff and fh are not kinetically/mechanically stable. Even if formed during the growth process, they should transform into the bcc-like stacking at considerable surface coverages.

Moreover, we computed the magnetic exchange interaction beyond first nearest neighbors, the DMI beyond first nearest neighbors and the magnetocrystalline anisotropy. 
We find considerable differences for all magnetic interactions depending on the Fe double-layer stacking. There are no general trends though. Therefore, every structure has to be analyzed individually.

In ff and fh we obtain a clockwise cycloidal spin spirals with periods of 1.6~nm and 1.1~nm, respectively, which are stabilized by the frustrated exchange and DMI, while the spin spirals in fb$^*$ and hf with wavelengths 3.4~nm and 3.0~nm are stabilized by the DMI.

The reduced symmetry arising from the surface Fe-layer in the bcc-like stackings (plane group c$m$ instead of p$3m1$ for the hexagonal close-packed structures) has fundamental consequences for the symmetry of the magnetic interactions. Not only the exchange interaction is affected, but also the DMI. We can explain this finding based on the density of states of the Fe $3d$ and Ir $5d$ electrons.
We show that an interfacial DMI may be influenced by a magnetic overlayer further from the interface.

This interaction between the surface Fe-layer and the Ir-substrate provides a new opportunity to tune the DMI by imposing different symmetries on the DMI. In the case of 2Fe/Ir (111), the 6-fold symmetric DMI present at the Ir-Fe interface can be modified by lowering the symmetry of the surface Fe-layer. The deposition of a bcc $5d$ transition metal on 2Fe/Ir (111) should create a large 2-fold symmetric DMI in the interface between Fe@vac and the $5d$ transition metal. The combination of two DMIs of 6-fold and 2-fold symmetry could enable the stabilization of topologically protected states with spatially varying chirality. This variation of chirality may allow the stabilization of higher order skyrmions (such as $S=-2$ or $S=-3$) in multilayer geometry.
 
\section*{Acknowledgements}
MD and BD are grateful for the computing time granted on the supercomputers of the North-German Supercomputing Alliance (HLRN) and Mogon at Johannes Gutenberg Universität Mainz [\onlinecite{MOGON}]. MD, BD and JS acknowledge financial support from the Alexander von Humboldt Foundation and the Transregional Collaborative Research Center (SFB/TRR) 173 SPIN+X. BD thanks the Deutsche Forschungsgemeinschaft (DFG) via the project DU 1489/2-1. SH thanks the European Unions Horizon 2020 research and innovation programme under grant agreement No. 665095 (FET-Open project MAGicSky). JS also acknowledges the  Grant Agency of the Czech Republic grant no. 14-37427G.

\bibliographystyle{apsrev4-1}

\bibliography{mylib}

\end{document}